%% file: mixing.tex
\documentclass[twocolumn, tighten, twocolappendix]{aastex631_mod}
\usepackage[T1]{fontenc}
\usepackage{microtype}
\usepackage{newtxtext}
\usepackage[varvw]{newtxmath}
\usepackage{mathtools}
\usepackage{enumitem}
\usepackage{booktabs}
\usepackage{multirow}
\usepackage{tikz}
\usepackage{bm}

\usetikzlibrary{shapes}
\hypersetup{pdfauthor={Errani et al.},
            pdftitle={},
            pdfkeywords={},
            bookmarksnumbered=true}

\newcommand{\rmx}{r_\mathrm{mx}}

\newcommand{\rh}{r_\mathrm{h}}

\newcommand{\diff}{\mathrm{d}}

\newcommand{\kpc}{\mathrm{kpc}}
\newcommand{\pc}{\mathrm{pc}}
\newcommand{\Gyr}{\mathrm{Gyr}}

\newcommand{\dex}{\mathrm{dex}}

\newcommand{\FeH}{\mathrm{[Fe/H]}}

\newcommand{\g}{\mathrm{g}}
\newcommand{\E}{\mathcal{E}}

\newcommand{\circleI}{\tikz[baseline={([yshift=-3pt]current bounding box.base)}]{\node[shape=circle,draw,inner sep=1.7pt, semithick, fill=starcol1] (char) { };}}
\newcommand{\circleII}{\tikz[baseline={([yshift=-3pt]current bounding box.base)}]{\node[shape=circle,draw,inner sep=1.7pt, semithick, fill=starcol2] (char) { };}}

\newcommand{\circleIII}{\tikz[baseline={([yshift=-3pt]current bounding box.base)}]{\node[shape=circle,draw,inner sep=1.7pt, semithick, fill=starcol3] (char) { };}}

\newcommand{\circleIV}{\tikz[baseline={([yshift=-3pt]current bounding box.base)}]{\node[shape=circle,draw,inner sep=1.7pt, semithick, fill=starcol4] (char) { };}}

\newcommand{\crossedoutcircle}{%
  \tikz[baseline={([yshift=-2.25pt]current bounding box.base)}]{
    \node[shape=circle, draw, inner sep=1.7pt, semithick, fill=none] (char) {};
    \node[draw=black, cross out, line width=0.7pt, line cap=round, scale=0.5] at (char) {};
  }%
}

\newcommand{\emptycircle}{%
  \tikz[baseline={([yshift=-2.25pt]current bounding box.base)}]{
    \node[shape=circle, draw, inner sep=1.7pt, semithick, fill=none] (char) {};
  }%
}

\definecolor{starcol1}{HTML}{d7191c}
\definecolor{starcol2}{HTML}{fdae61}
\definecolor{starcol3}{HTML}{abd9e9}
\definecolor{starcol4}{HTML}{2c7bb6}

\graphicspath{{./}{./Figures/}}

\begin{document}
\shorttitle{Impulsive mixing of stellar populations}
\shortauthors{Errani et al.}
\title{Impulsive mixing of stellar populations in dwarf spheroidal galaxies}

\author{Rapha\"el Errani}
\affiliation{McWilliams Center for Cosmology and Astrophysics, Department of Physics, Carnegie Mellon University, Pittsburgh, PA 15213, USA}
\email{errani@cmu.edu}

\author{Matthew G. Walker}
\affiliation{McWilliams Center for Cosmology and Astrophysics, Department of Physics, Carnegie Mellon University, Pittsburgh, PA 15213, USA}

\author{Simon Rozier}
\affiliation{School of Mathematics and Maxwell Institute for Mathematical Sciences, University of Edinburgh, Kings Buildings, Edinburgh, EH9 3FD, UK}

\author{Jorge Pe\~narrubia}
\affiliation{Institute for Astronomy, University of Edinburgh, Royal Observatory, Blackford Hill, Edinburgh EH9 3HJ, UK}

\author{Julio F. Navarro}
\affiliation{Department of Physics and Astronomy, University of Victoria, Victoria, BC, V8P 5C2, Canada}

\received{2025 February 25}
\revised{2025 July 7}
\accepted{2025 August 6}

\begin{abstract}
We study the response of mono-energetic stellar populations with initially isotropic kinematics to impulsive and adiabatic changes to an underlying dark matter potential. Half-light radii expand and velocity dispersions decrease as enclosed dark matter is removed. The details of this expansion and cooling depend on the time scale on which the underlying potential changes. In the adiabatic regime, the product of half-light radius and average velocity dispersion is conserved. We show that the stellar populations maintain centrally isotropic kinematics throughout their adiabatic evolution, and their densities can be approximated by a family of analytical radial profiles. Metallicity gradients within the galaxy flatten as dark matter is slowly removed. In the case of strong impulsive perturbations, stellar populations develop power-law-like density tails with radially biased kinematics. We show that the distribution of stellar binding energies within the dark matter halo substantially widens after an impulsive perturbation, no matter the sign of the perturbation. This allows initially energetically separated stellar populations to mix, to the extent that previously chemo-dynamically distinct populations may masquerade as a single population with large metallicity and energy spread. Finally, we show that in response to an impulsive perturbation, stellar populations that are deeply embedded in cored dark matter halos undergo a series of damped oscillations before reaching a virialized equilibrium state, driven by inefficient phase mixing in the harmonic potentials of cored halos. This slow return to equilibrium adds substantial systematic uncertainty to dynamical masses estimated from Jeans modeling or the virial theorem. 
\end{abstract}
\keywords{Cold dark matter (265); Dwarf spheroidal galaxies (420); Galaxy dynamics (591); $N$-body simulations (1083); Perturbation methods (1215)}

\input{1_introduction}

\input{2_setup}

\input{3_adiabatic}

\input{4_impulsive}

\input{5_summary}

\section*{Acknowledgements}
RE and MW acknowledge support from the National Science Foundation (NSF) grant AST-2206046. SR acknowledges support from the Royal Society (Newton International Fellowship NIF\textbackslash R1\textbackslash 221850). JFN would like to acknowledge the hospitality of the Max-Planck Institute for Astrophysics and of the Donostia International Physics Centre during the completion of this manuscript. Support for program JWST-AR-02352.001-A was provided by NASA through a grant from the Space Telescope Science Institute, which is operated by the Association of Universities for Research in Astronomy, Inc., under NASA contract NAS 5-03127. This material is based upon work supported by the National Aeronautics and Space Administration under Grant/Agreement No. 80NSSC24K0084 as part of the Roman Large Wide Field Science program funded through ROSES call NNH22ZDA001N-ROMAN.

\appendix
\input{Appendix}

\newpage

\bibliographystyle{aasjournal-hyperref}
\bibliography{mixing}

\label{lastpage}
\end{document}

%% file: 1_introduction.tex
\section{Introduction}
Dwarf spheroidal (dSph) galaxies are considered to be among the most dark matter-dominated objects in the Universe. With half-light radii spanning sizes of $\sim \kpc$ \citep{Torrealba2016,Torrealba2019} down to $\sim 30\,\pc$ \citep{Belokurov2009, Koposov2015}, and potentially even smaller \citep{Smith2024, Simon2024}, these systems probe the distribution of dark matter on subgalactic scales, where dark matter particle properties may play a crucial role in setting the dark matter densities \citep{Nadler2021, Nadler2024}. 
In particular at the low-luminosity end, dSph galaxies are theorized to offer a particularly pristine glimpse at the distribution of dark matter, as stellar feedback in these systems is considered insufficiently violent to substantially alter their surrounding dark matter halos \citep{Penarrubia2012,DiCintio2014, Onorbe2015}.

dSph galaxies exhibit a rich diversity in structural, kinematic, and chemical parameters \citep[for reviews, see][]{McConnachie2012,Simon2019Review,Battaglia2022}. While the brighter dSph galaxies contain multiple chemo-dynamically distinguishable stellar populations with limited metallicity dispersion \citep[e.g.,][]{Tolstoy2004, Battaglia2006, Pace2020}, several Milky Way ultra-faint galaxies appear to consist of a single stellar population with a relatively large metallicity spread \citep[e.g.,][]{Kirby2013, Frebel2014, Koch2014, Walker2016_Grus1, Kirby2017}; i.e., for those systems, there is no obvious trend with metallicity in the spatial distribution or the kinematics of stars. Developing an understanding of the cause for this diversity is complicated by the multitude of external and internal processes that dSphs are subject to, in particular once accreted onto the Milky Way and subject to its tidal field. 

The detailed tidal evolution of dSph galaxies has been shown to depend on the mass profile of the underlying dark matter halo \citep{Penarrubia2008, Penarrubia2010, EPT15, EPLG17}, its flattening \citep{Sanders2018}, and the distribution of the stellar binding energies in the dark matter potential \citep{ENIP2022}. In addition, the evolution of the underlying dark matter halo will depend on its internal kinematics, and on any potential evolution driven by dark matter self-interactions (such as evaporation and core collapse, see, e.g., \citealt{Despali2019, Zeng2022}), or by outflows triggered by stellar feedback \citep[e.g.][]{Pontzen2012, Freundlich2020, Li2023}. Instead of focusing on the details of each of these processes, some intuition may be gained by categorizing the types of perturbation that may affect a stellar system. 

Gravitational perturbations can be broadly categorized by their amplitude and by the time scale on which they act, relative to the corresponding internal (physical and time) scales of the perturbed system. \emph{Linear} perturbations, in contrast to \emph{nonlinear} ones, are small in amplitude with respect to the gravitational binding energy of the stellar system in question. \emph{Adiabatic} perturbations, in contrast to \emph{impulsive} ones, act on time scales that are much longer than the dynamical time scale of the perturbed system. \emph{Resonant} perturbations have a characteristic time scale that matches an internal time scale of the perturbed system.

The response of stellar (and dark matter) distributions to slow perturbations has been studied using adiabatic invariants: \citet{Young1980} used the combined conservation of angular momentum and radial action to describe the response of a star cluster to a growing black hole; \citet{Barnes1984} studied the response of spheroids to a growing disc potential; \citet{Blumenthal1986} and \citet{Gnedin2004} described the adiabatic contraction of a dark matter halo due to accretion of baryons (see also \citealt{Sellwood2005} and \citealt{Katz2014}, who applied the \citealt{Young1980} algorithm). More recently, \citet{Stucker2023} used the \citet{Young1980} method to describe the tidal evolution of a dark matter halo in the adiabatic limit. For fast perturbations, the impulsive approximation was applied by \citet{Aguilar1985} and \citet{Banik2021_heating} for the study of the tidal interaction between spherical galaxies; \citet{GnedinOstriker1999} and \citet{GnedinHernquist1999} described the response of stellar systems to tidal shocks, and \citet{TaylorBabul2001,Drakos2020,BensonDu2022} applied the impulse approximation to model the tidal evolution of dark matter subhalos. For the regime of linear perturbations, \citet{RozierErrani2024} used the \citet{Kalnajs1977} matrix method of linear response theory to model the self-consistent response of gravitating systems to a perturbation, showing that adiabatic and impulsive perturbations result in the same final equilibrium state for linear perturbations. 

In the present work, we study the conditions necessary for initially chemo-dynamically distinct stellar populations to mix in response to a perturbation. We model the response of mono-energetic stellar tracer populations to linear and nonlinear perturbations, both in the adiabatic and in the impulsive regime, making use of controlled $N$-body simulations. This is useful because stellar populations in dSph galaxies may be thought of as a combination of multiple populations of massless tracers with spherical density profiles, mono-energetic distribution functions, and isotropic kinematics. Stellar profiles with complicated energy distributions may be constructed by linear superposition of the mono-energetic results presented here. We discuss the response of mono-energetic stellar tracers to perturbations in dark matter halos with centrally-divergent density cusps, and central constant-density cores.

The paper is structured as follows. 
In Sec.~\ref{sec:NumericalSetup}, we introduce the numerical setup used in the present study, and derive a family of stellar density profiles with mono-energetic distribution functions, embedded in cuspy and cored dark matter halos. 
In Sec.~\ref{sec:Adiabatic}, we consider the case of adiabatic perturbations, and show that initially energetically separated stellar populations remain separated. 
In Sec.~\ref{sec:Impulsive}, we turn our attention to impulsive perturbations, where stellar populations may mix energetically and develop radially biased kinematics.
For stellar tracers embedded in cored halos, we show that inefficient phase mixing results in long-lived oscillations in response to an impulsive perturbation. 
Finally, in Sec.~\ref{sec:Summary}, we summarize our main findings.

%% file: 2_setup.tex
\section{Numerical setup}
\label{sec:NumericalSetup}
We discuss here the numerical setup adopted to study the response of mono-energetic stellar tracer populations to impulsive and adiabatic changes to an underlying dark matter potential. As an illustration, we begin by showing how an arbitrary stellar distribution function of a spherical stellar component with isotropic kinematics can be approximated by a sum of mono-energetic profiles. 

\subsection{Distribution function for spherical stellar systems}
Consider a star on an orbit in a spherical dark matter potential. As an example, we choose a \citet{hernquist1990} potential, which in the 
\citet{dehnen1993} parameterization corresponds to the case of $\gamma=1$, with density profile
\begin{equation}
 \label{eq:RhoDehnenGamma}
 \rho_\gamma(r) = \frac{(3-\gamma) M}{4\pi} \frac{a}{r^\gamma (r+a)^{4-\gamma}}~.
\end{equation}
In the above equation, $\gamma$ denotes the central logarithmic slope, $M$ is the total dark matter mass, and $a$ is the dark matter scale radius. A value of $\gamma=1$ corresponds to a centrally-divergent cuspy profile, whereas $\gamma=0$ produces a profile with a central constant-density core. 

For a potential with a finite central value $\Phi_0$ and boundary condition $\Phi \rightarrow 0$ at infinity, we define the dimensionless energy $\E$ of a star located at radius $r$ with velocity $v$ as
\begin{equation}
 \label{eq:EnergyDef}
 \E \equiv 1 - E / \Phi_0 
\end{equation}
where $E=v^2/2 + \Phi(r)$. A stationary star located at $r=0$ has energy $\E=0$ (the ground state), and $\E=1$ denotes the boundary between bound and unbound orbits.
Analogously, we define the (dimensionless) potential as
\begin{equation}
 \label{eq:PotentialDef}
 \psi(r) \equiv 1 - \Phi(r) / \Phi_0
\end{equation}
i.e. $\psi=0$ corresponds to the potential minimum, and $\psi = 1$ is the boundary condition at infinity. Using these definitions, $\E = u^2/2 + \psi(r)$ with $u\equiv v/\sqrt{-\Phi_0}$. 

\begin{figure}[tb]
  \centering
  \includegraphics[width=8.5cm]{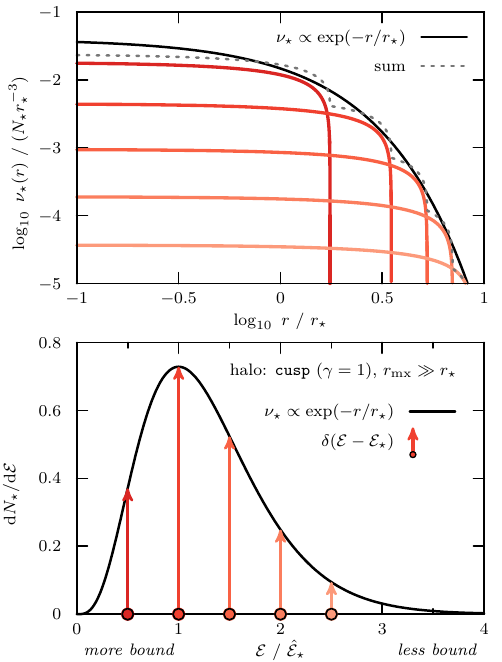}
  \caption{Top panel: 3D exponential number density $\nu_\star(r)$, Eq.~\ref{eq:3Dexp_rho}, of a stellar tracer with $N_\star$ stars and scale radius $r_\star\approx\rh/2.67$ (black solid curve), as well as an approximation to this profile consisting of the sum of five mono-energetic stellar distributions. Red curves show the individual mono-energetic profiles, Eq.~\ref{eq:monoenergetic_rho}, and a dashed curve shows their sum, with each component weighted by $\left. \diff N_\star / \diff \E \right|_{\E = \E_\star}$. Bottom panel: Differential energy distribution $\diff N_\star / \diff \E$, Eq.~\ref{eq:energydist}, of an exponential stellar profile embedded deeply in the density cusp of a Hernquist dark matter halo, Eq.~\ref{eq:Hernquist_pot}. Energies are expressed in units of the energy $\hat \E_\star$ where $\diff N_\star / \diff \E$ peaks, and the energy distribution is normalized here so that $\int_0^1 \left(\diff N_\star/\diff \E\right) \diff \E = 1$. The five $\delta$-functions corresponding to the mono-energetic profiles shown in the top panel are depicted as arrows. }
  \label{fig:DeltaSumIllustration}
\end{figure}

The dimensionless potential of a (cuspy, $\gamma=1$) Hernquist profile can be written as
\begin{equation}
\label{eq:Hernquist_pot}
 \psi_{\gamma=1}(r) = \left. r \middle/ (r+a) \right. ~.
\end{equation}
As an example, we embed a massless stellar tracer population in this potential, with a (3D) exponential number density profile, 
\begin{equation}
\label{eq:3Dexp_rho}
 \nu_\star(r) = N_\star~ \frac{\exp(-r/r_\star)}{8\pi r_\star^3} ~.
\end{equation}
In Eq.~\ref{eq:3Dexp_rho} we denote by $N_\star$ the total number of stars, and $r_\star$ is a scale radius. We refer to the radius that encloses half of the stars as the half-light radius, which, for an exponential profile, equals $ \rh \approx 2.67\,r_\star$. The number density profile is shown in the top panel of Fig.~\ref{fig:DeltaSumIllustration} (see also Fig.~12 in \citealt{EINPW2024} for a comparison of this profile against observed stellar densities of Milky Way dSphs).

Under the assumptions of spherical symmetry and isotropic kinematics, the distribution function $f_\star(\E)$, which describes the number of stars $N_\star$ per phase space volume element $\diff \Omega = \diff^3 r \diff^3 v$, can be computed from Eddington inversion \citep[see e.g. Eq.~4-140b in][]{BT87}. Using the definitions of Eq.~\ref{eq:EnergyDef} and \ref{eq:PotentialDef}, 
\begin{equation}
 f_\star(\E) \equiv \frac{\diff N_\star}{\diff \Omega} = \frac{1}{\sqrt{8} \pi^2  (-\Phi_0)^{3/2}  } \int\limits_{\E}^1 \frac{\diff^2 \nu_\star}{\diff \psi^2} \frac{\diff \psi}{\sqrt{\psi - \E}} ~,
\end{equation}
where we have assumed that the stellar number density $\nu_\star(r)$ vanishes sufficiently fast toward infinity so that $\left.{\diff \nu_\star}/{\diff \psi} \right|_{\psi=1} = 0$. For an exponential stellar tracer that is deeply embedded within the power-law cusp of the Hernquist halo, $r_\star \ll a$, the potential may be approximated by $ \psi(r) \approx {r} / {a}$ and $ {\diff^2 \nu_\star} / {\diff \psi^2} \approx \nu_\star(\psi) a^2 / r_\star^2 $.
For these assumptions, the distribution function has a simple analytical form,
\begin{equation}
 f_\star(\E) \stackrel{\E \ll 1}{\approx} \frac{N_\star}{16\sqrt{2\,\pi^5}} ~ \sqrt{ \frac{a^3}{r_\star^{9} (-\Phi_0)^3} } ~  \exp\left( - \frac{ \E}{ r_\star/a  } \right)~.
\end{equation}
In the density cusp of a Hernquist halo, the phase space volume~$\diff \Omega$ accessible per (dimensionless) energy bin $\diff \E$ becomes
\begin{eqnarray}
 g(\E) \equiv \frac{\diff \Omega}{\diff \E} &=&  (-\Phi_0)^{3/2} \int_0^{\psi^{-1}(\E)} \!\! \sqrt{2(\E-\psi)} ~ r^2 ~\diff r \\
      &\stackrel{\E \ll 1}{\approx}& \frac{16 \sqrt{2}}{105} ~a^3~ (-\Phi_0)^{3/2}~ \E^{7/2} ~.
\end{eqnarray}
The differential energy distribution of a deeply embedded stellar tracer ($r_\star \ll a$)
\begin{equation}
\label{eq:energydist}
\diff N_\star / \diff \E = 16\,\pi^2 f_\star(\E) g(\E) \stackrel{\E \ll 1}{\propto}   \E^{7/2}  \exp\left( - \frac{ \E}{ r_\star/a  } \right)
\end{equation}
peaks at an energy of $\hat \E_\star = (7/2) (r_\star/a)$. Note that the functional form of Eq.~\ref{eq:energydist} is identical to the empirically motivated form proposed in \citet{ENIP2022} to parameterize stellar energy distributions in LCDM. 
The differential energy distribution Eq.~\ref{eq:energydist} is shown as a black curve in the bottom panel of Fig.~\ref{fig:DeltaSumIllustration}. 

\subsection{Decomposition into mono-energetic profiles}
We will now approximate the energy distribution of Eq.~\ref{eq:energydist} by a sum of delta functions, and analogously, approximate the density profile of Eq.~\ref{eq:3Dexp_rho} by a sum of mono-energetic profiles. For this, we aim to derive the functional form for the (stellar) number density corresponding to a mono-energetic distribution function in a spherical (dark matter) potential (following the steps outlined in Appendix~C of \citealt{EINPW2024}). We consider a mono-energetic distribution function 
\begin{equation}
 \label{eq:mono_energetic_DF}
 f_\star (\E) \propto \delta(\E-\E_\star)
\end{equation}
centred around some energy $\E_\star$. We compute the number density profile by integrating the distribution function over all velocities $v$, which for $\psi(r) \leq \E_\star$ gives
\begin{equation}
 \nu_\star(r) \propto \int\limits_{\text{all }v} v^2 ~\delta \left( \E - \E_\star \right) \, \diff v ~\propto~ \sqrt{ \E_\star - \psi(r) } ~.
\end{equation}
This profile above has a finite central value for all potentials with finite central escape velocity $v_\mathrm{esc}=\sqrt{-2\Phi_0}$, i.e., the profile is cored. 
Normalized to its central value,
\begin{equation}
 \label{eq:monoenergetic_rho}
 \nu_\star(r)/\nu_\star(0) =
 \begin{dcases}
   \sqrt{ 1 - \psi(r)/\E_\star } ~  &\text{if } \psi(r) \leq \E_\star\\
   0~ &\text{otherwise.}
 \end{dcases}
\end{equation}
The profile is sharply truncated at a radius $r_\mathrm{t}$ where $\psi(r_\mathrm{t}) = \E_\star$.
We show the profile of Eq.~\ref{eq:monoenergetic_rho} in the top panel of Fig.~\ref{fig:DeltaSumIllustration} for five example values of $\E_\star$, with the corresponding delta function energy distributions shown in the bottom panel.

The velocity dispersion profile can be easily computed noting that for mono-energetic distributions, all particles at a given radius $r$ have the same kinetic energy. 
As the distribution function Eq.~\ref{eq:mono_energetic_DF} is a function of energy alone, the kinematics are isotropic and 
\begin{equation}
\label{eq:StellarSigma}
 \sigma^2(r) = -2\Phi_0~ \left[ \E_\star - \psi(r) \right]~.
\end{equation}
Hence,
\begin{equation}
  \sigma(r) / \sigma(0) =  \nu_\star(r)/\nu_\star(0)
\end{equation}
for a central value of $\sigma(0) = \sqrt{- 2\Phi_0 \E_\star }$.

Density profiles are additive, and velocity dispersion profiles are additive in quadrature, i.e. the profiles corresponding to a more complicated stellar distribution function can be approximated by the linear superposition of mono-energetic profiles. The gray dashed curve in the top panel of Fig.~\ref{fig:DeltaSumIllustration} shows such an approximation by a sum of equally spaced delta functions, each mono-energetic profile of energy $\E_\star$ being weighted by $\left. \diff N_\star / \diff \E \right|_{\E = \E_\star}$.  

\subsection{Stellar and dark matter profiles considered in this work}
\label{sec:stellar_profiles}

\begin{figure}[tb]
  \centering
  \includegraphics[width=8.5cm]{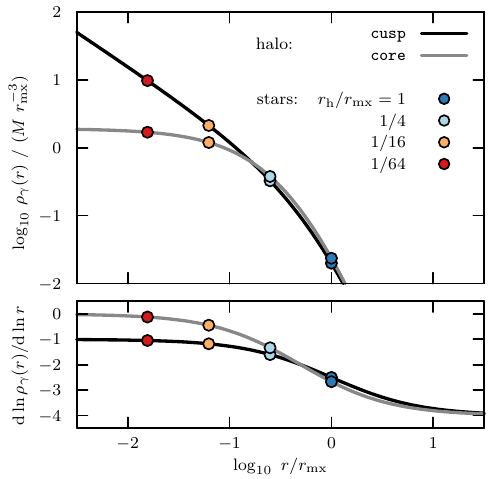}
  \caption{Top panel: cuspy (black, Eq.\ref{eq:RhoDehnenGamma} with $\gamma=1$) and cored (gray, $\gamma=0$) dark matter halo density profiles considered in this study. The half-light radii of four mono-energetic stellar tracers with $\rh / \rmx =1$, $1/4$, $1/16$ and $1/64$ are indicated by filled circles in dark blue, light blue, orange and red, respectively (see Tab.~\ref{tab:parameter_overview} for the corresponding energies).  Bottom panel: 3D logarithmic slope of the dark matter density profiles shown above.}
  \label{fig:DMdensities}
\end{figure}

In this work, we study the evolution of stellar tracer populations embedded in a cuspy ($\gamma=1$) and a cored ($\gamma=0$) dark matter halo parameterized by Eq.~\ref{eq:RhoDehnenGamma}. The two dark matter density profiles are shown in the top panel of Fig.~\ref{fig:DMdensities}. 
The bottom panel shows the logarithmic slope
\begin{equation}
\label{eq:SlopeDehnenGamma}
 \left. \diff \ln  \rho_\gamma(r) \middle/ \diff \ln r \right. = -\gamma - (4-\gamma) ~r / (r+a)~,
\end{equation}
which, for $r \ll a$, approaches $-\gamma$, whereas for $r \gg a$, the logarithmic slope of the density profile asymptotically approaches $-4$. 
For the cuspy ($\gamma=1$) profile, the circular velocity curve $V_\mathrm{c}(r) = \sqrt{GM(<r)/r}$ peaks at a radius $r_\mathrm{mx} = a$ with circular velocity $V_\mathrm{mx} = \sqrt{-\Phi_0}/2$, and $\Phi_0 = - GM/a$. For the cored profile ($\gamma=0$), we have $r_\mathrm{mx} = 2a$ with a peak circular velocity of $V_\mathrm{mx} = \sqrt{-8 \Phi_0 / 27 }$ and $\Phi_0=-GM/(2a)$. For both models, $\diff \ln \rho / \diff \ln r = -2$ at a radius of $2r_\mathrm{mx}$.

\begin{table}[tb]
    \centering
    \caption{Initial conditions for four mono-energetic stellar tracers (Eq.~\ref{eq:monoenergetic_rho}) embedded in a cuspy (Eq.~\ref{eq:Hernquist_pot}) and a cored (Eq.~\ref{eq:coredPsi}) dark matter halo. Listed are initial half-light radii $r_\mathrm{h0}$, normalized by the radius of maximum circular velocity $r_\mathrm{mx0}$ of the underlying halo. Stellar binding energies $\E_\star = 1 - E_\star/\Phi_0$ as well as the average velocity dispersion of all stars $\langle \sigma^2_0 \rangle$ are expressed in units of the potential minimum $\Phi_0 \equiv \Phi(r=0)$.}
    \begin{tabular}{l@{\hskip 0.2cm}l@{\hskip 0.9cm}l@{\hskip 0.2cm}l@{\hskip 0.5cm}l@{\hskip 0.2cm}l}
    \toprule
                  & &\multicolumn{2}{l}{$~~$\texttt{cusp} ($\gamma=1$)}&\multicolumn{2}{l}{$~~$\texttt{core} ($\gamma=0$)} \\[0.1cm]                     
                  & $\displaystyle \frac{r_\mathrm{h0}}{r_\mathrm{mx0}}$ & $\log_{10}\E_\star$ & $\displaystyle \log_{10} \frac{ \langle \sigma^2_0 \rangle}{-\Phi_0}$ & $\log_{10}\E_\star$ & $\displaystyle \log_{10} \frac{ \langle \sigma^2_0 \rangle}{-\Phi_0}$  \\[0.2cm]  \midrule
       \circleIV  &   $1$             & -0.22                           & -0.62 & -0.25  &-0.55  \\  
       \circleIII &   $1/4$           & -0.57                           & -0.82 & -0.76  &-0.86 \\  
       \circleII  &   $1/16$          & -1.08                           & -1.28 & -1.64  &-1.67 \\  
       \circleI   &   $1/64$          & -1.66                           & -1.84 & -2.75  &-2.75 \\  
       \bottomrule
    \end{tabular}
    \label{tab:parameter_overview}
\end{table}

We embed four mono-energetic stellar profiles within these halos, with half-light radii of $\rh/\rmx = 1$, $1/4$, $1/16$ and $1/64$. The half-light radii are marked in Fig.\ref{fig:DMdensities} using filled circles in dark blue, light blue, orange and red, respectively. The corresponding energies are listed in Tab.~\ref{tab:parameter_overview}.

\subsubsection{Cuspy Hernquist ($\gamma=1$) halo}
\label{sec:SetupCuspyStars}
\begin{figure}[tb]
  \centering
  \includegraphics[width=8.5cm]{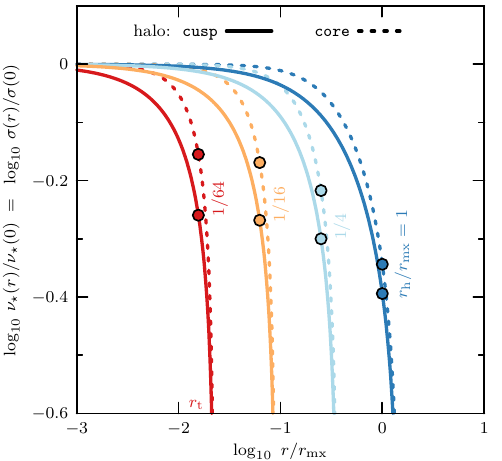}
  \caption{Mono-energetic stellar number density profiles $\nu_\star(r)$ (Eq.~\ref{eq:monoenergetic_rho}) with energies as listed in Tab.~\ref{tab:parameter_overview}, with half-light radii of $\rh / \rmx =1$, $1/4$, $1/16$ and $1/64$. Stellar profiles embedded in a cuspy (Eq.~\ref{eq:StellarCuspy}) and a cored (Eq.~\ref{eq:StellarCored}) dark matter halo are shown using solid and dashed lines, respectively. Filled circles indicate the locations of the respective half-light radii. The profiles are sharply truncated at a radius $r_\mathrm{t}$. Note that for these mono-energetic energy distributions, the relative density and velocity dispersion profiles coincide.  }
  \label{fig:StellarDensitiesCuspCore}
\end{figure}

We now derive the functional form of the stellar density profiles for the example cases of the cuspy and cored dark matter halos of Eq.~\ref{eq:RhoDehnenGamma} shown in Fig.~\ref{fig:DMdensities}. 
Using the same notation as above, for a cuspy Hernquist potential as in Eq.~\ref{eq:Hernquist_pot}, 
\begin{equation}
 \label{eq:StellarCuspy}
 \nu_\star(r)/\nu_\star(0) = \sqrt{ 1 -  r / \left[ (r+a) \E_\star  \right] }~,
\end{equation}
truncated beyond a radius
\begin{equation}
 r_\mathrm{t}/a = \E_\star / \left ( {1-\E_\star} \right) ~.
 \label{eq:HernquistStellarTruncation}
\end{equation}
The radius $r_{-2}$, where $\diff N_\star / \diff r = 4\pi r^2 \nu_\star(r)$ peaks and most stars are located, has the expression
\begin{equation}
 r_{-2}/a =  \left[ \E_\star - 5/8 + \sqrt{ (5/8)^2 - \E_\star/4} \right] \left(1-\E_\star\right)^{-1}.
\end{equation}
In the limit of a stellar tracer embedded deeply inside the density cusp, i.e. $r_{-2} \ll 1$, $r_{-2} \sim 4\,\E_\star/5$.
For the same limit, 
\begin{equation}
 \nu_\star(0) = \frac{105}{64\pi}~N_\star~r_\mathrm{t}^{-3} = \frac{21}{25\pi}~ N_\star~ r_{-2}^{-3}~.
\end{equation}
The stellar number density profiles embedded in a cuspy halo are shown using solid curves in Fig.~\ref{fig:StellarDensitiesCuspCore}. The profiles are centrally flat, and sharply truncated at the radius $r_\mathrm{t}$.

\subsubsection{Cored Dehnen ($\gamma=0$) halo}
\label{sec:SetupCoredStars}
For the case of a dark matter halo with a centrally-constant density, parameterized by a \citet{dehnen1993} profile (Eq.~\ref{eq:RhoDehnenGamma}) with $\gamma=0$, 
\begin{equation}
 \label{eq:coredPsi}
 \psi(r) =  {r^2} / \left ( {r+a} \right)^2 
\end{equation}
and 
\begin{equation}
 \label{eq:StellarCored}
 \nu_\star(r)/\nu_\star(0) = \sqrt{ 1 -  {r^2} / \left[  (r+a)^2 \E_\star \right]  } ~,
\end{equation}
truncated beyond a radius
\begin{equation}
 r_\mathrm{t}/a = \left. {\sqrt{\E_\star}}  \middle/ \left(   {1-\sqrt{\E_\star}}  \right) \right.~.
 \label{eq:DehnenStellarTruncation}
\end{equation}
The radius $r_{-2}$ where most stars are located is the solution to $\E_\star = \left(r_{-2}\right)^2 \left(r_{-2} + 3a/2\right) \left(r_{-2}+a\right)^{-3} $, i.e. for a stellar system embedded deeply inside the constant-density core, $r_{-2}/a \ll 1$, $r_{-2}/a \sim \sqrt{2\,\E_\star/3}$.
For the same limit, 
\begin{equation}
 \nu_\star(0) = \frac{4}{\pi^2}~N_\star~r_\mathrm{t}^{-3} = \frac{\sqrt{128}}{\sqrt{27}\,\pi^2}~ N_\star~ r_{-2}^{-3}~.
\end{equation}

The stellar density profiles embedded in a cored halo are shown using dashed curves in Fig.~\ref{fig:StellarDensitiesCuspCore}. 

\begin{figure}[tb]
 \centering
 \includegraphics[width=8.5cm]{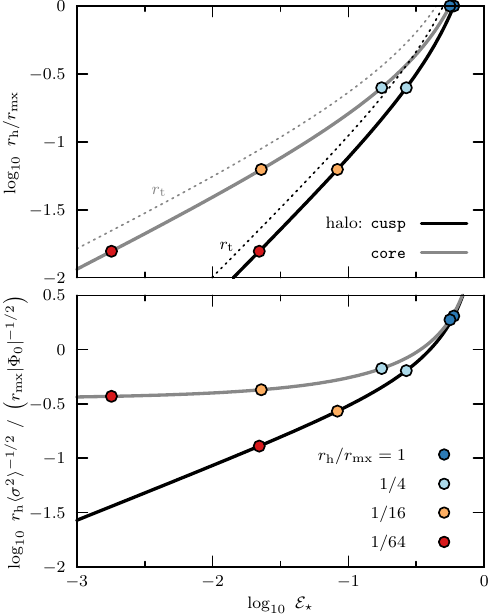} 
   \caption{Top panel: Relation between stellar binding energy $\E_\star$ and half-light radius $\rh$ for mono-energetic stellar tracers embedded in a cuspy (black solid curve) and a cored (gray solid curve) potential. The values corresponding to the models of Fig.~\ref{fig:StellarDensitiesCuspCore} are shown as filled circles. The dashed curves show the truncation radii $r_\mathrm{t}$ beyond which there is no stellar density (Eq.~ \ref{eq:HernquistStellarTruncation}, \ref{eq:DehnenStellarTruncation}). The energetic spacing of stellar populations with matching half-light radii differs between models embedded deeply in a cuspy and cored halo. Bottom panel: Crossing time $T_\mathrm{cross} = \rh \langle \sigma^2 \rangle^{1/2}$ as a function of energy $\E_\star$. Stellar populations that are deeply embedded in the density core all have virtually identical crossing times, regardless of their binding energy.}
   \label{fig:E_vs_rh_relation}
\end{figure}

To summarize, we show the general relation between the energy $\E_\star$ of a stellar delta function and the corresponding half-light radius $\rh$ of the corresponding mono-energetic stellar population in the top panel of Fig.~\ref{fig:E_vs_rh_relation}. The energies and half-light radii of the four mono-energetic profiles considered in this study are highlighted using filled circles. Note that the slope $\diff \rh / \diff \E_\star$ differs between the cuspy and the cored model for the case of deeply embedded stellar populations, i.e., the spacing in energy of stellar populations with similar half-light radii differs between models embedded in a cuspy (black solid curve) and a cored (gray solid curve) halo. For reference, the truncation radius $r_t$ (Eq.~\ref{eq:HernquistStellarTruncation},~\ref{eq:DehnenStellarTruncation}) is shown as a dashed curve in the same figure. For future reference, the bottom panel of Fig.~\ref{fig:E_vs_rh_relation} shows the relation between crossing time $T_\mathrm{cross} = \rh \langle \sigma^2 \rangle^{-1/2}$ and binding energy, where by $\langle \sigma^2 \rangle^{1/2}$ we denote the average velocity dispersion of all stars.

\subsection{N-body models}
To study the response of the stellar distributions to changes of the underlying potential, we build equilibrium $N$-body realizations of the mono-energetic stellar distributions described in Sec.~\ref{sec:SetupCuspyStars} and \ref{sec:SetupCoredStars}. Stars are assumed to be mass-less, i.e., their kinematics are determined solely by the underlying (dark matter) potential. Each $N$-body realization consists of $10^4$ particles for the simulations with adiabatically evolving background potential (Sec.~\ref{sec:Adiabatic}), and of $10^5$ particles for the runs with impulsive perturbations (Sec.~\ref{sec:Impulsive} and \ref{sec:deepening}). Selected runs use a higher $N$-body particle number as stated in the text. The background potential is analytical and given by Eq.~\ref{eq:Hernquist_pot} for the cuspy potential and by Eq.~\ref{eq:coredPsi} for the cored one. The time evolution of the background potential will be detailed in Sec.~\ref{sec:Adiabatic} for the adiabatic case and in Sec.~\ref{sec:Impulsive} for the impulsive one. The orbit integration of the $N$-body particles is carried out using a classical Runge-Kutta scheme of fourth order with an adaptive time step criterion, ensuring that each circular orbit is resolved by at least 100 time steps.

%% file: 3_adiabatic.tex
\section{Adiabatic perturbations}
\label{sec:Adiabatic}

\begin{figure*}[tb]
 \centering
 \includegraphics[width=18cm]{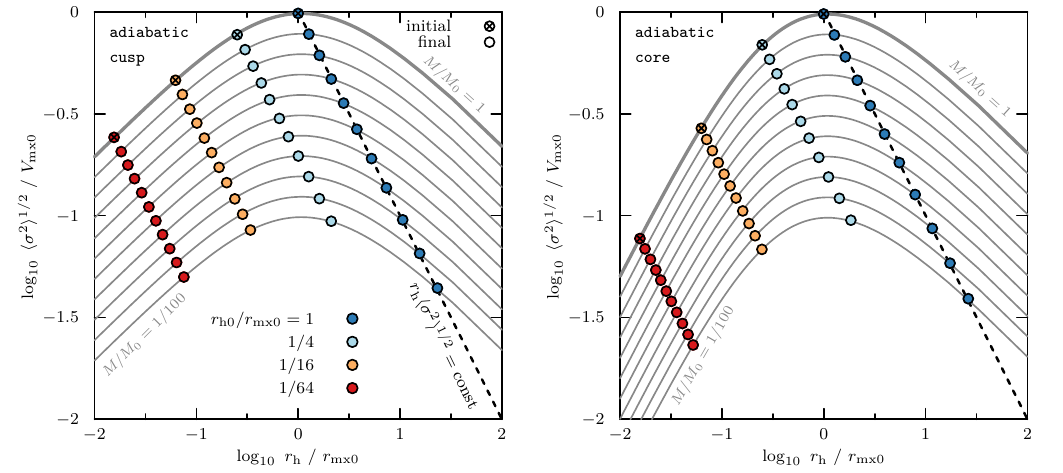} 
\caption{Initial (\crossedoutcircle{}) and final (\emptycircle{}) equilibrium configurations of stellar tracer populations in the numerical experiments with an adiabatically evolving dark matter halo. Shown are half-light radii $\rh$ and average velocity dispersions $\langle \sigma^2 \rangle^{1/2}$ of four stellar tracers with mono-energetic initial conditions, embedded in a cuspy ($\gamma=1$ in Eq.~\ref{eq:RhoDehnenGamma}, left panel) and a cored ($\gamma=0$, right panel) dark matter halo. Stellar tracers with initial half-light radii $r_\mathrm{h0} / r_\mathrm{mx0} = 1$, $1/4$, $1/16$ and $1/64$ are shown in dark blue, light blue, orange and red, respectively. All initial conditions lie along the thick gray curve labeled $M/M_0=1$. In each simulation, the mass sourcing the underlying dark matter potential is lowered slowly by $-0.2\,\dex$, $-0.4\,\dex$, $\dots$, $-2\,\dex$. Solid gray curves are computed from Eq.~\ref{eq:LumAveragedDispersion}, assuming mono-energetic distribution functions with isotropic kinematics. All equilibrium configurations fall along lines of $\rh \langle\sigma^2\rangle^{1/2}=\text{const}$. }
\label{fig:AdiabaticRhSigmaOverview}
\end{figure*}

\begin{figure}[tb]
 \centering 
 \includegraphics[width=8.5cm]{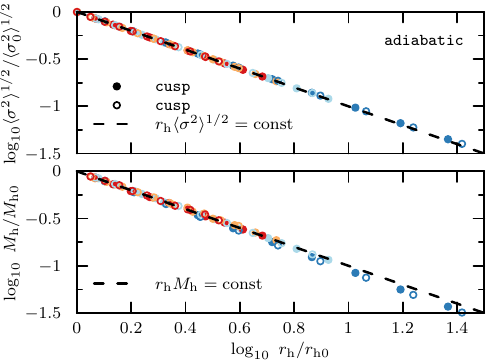} 
\caption{Top panel: Average velocity dispersion $\langle \sigma^2 \rangle^{1/2}$ as a function of half-light radius $\rh$, relative to their respective initial values. The same simulation snapshots are shown as in Fig.~\ref{fig:AdiabaticRhSigmaOverview}, each point corresponding to an equilibrium configuration reached after an adiabatic change to the underlying potential. By plotting the relative evolution of structural parameters, the systematics of Fig.~\ref{fig:AdiabaticRhSigmaOverview} can be collapsed onto a single curve. Bottom panel: as above, but showing the remnant (dark matter) mass $M_\mathrm{h} \equiv M(<\rh)$ enclosed within the current half-light radius.}
\label{fig:AdiabaticCollapsed}
\end{figure}

\subsection{Adiabatic evolution of size and kinematics}
\label{sec:bulk_structure_adiabatic}
We will now turn our attention to the evolution of stellar populations subject to slow changes to the underlying potential. Specifically, we study how the four mono-energetic stellar distributions introduced in Sec.~\ref{sec:stellar_profiles} respond to slow changes to the (dark matter) mass sourcing the underlying potential. We study the evolution separately for stars embedded in a cuspy and a cored dark matter halo. We slowly decrease the total mass~$M$ of each dark matter halo, while keeping its radius of peak circular velocity $\rmx$ unaltered (see Eq.~\ref{eq:RhoDehnenGamma} for definition). To ensure that the change of the underlying potential is perceived as adiabatic by each stellar population, we decrease the mass~$M$ exponentially with a decay constant that is at least 10 times longer than the period of a circular orbit at the (final) half-light radius in the new equilibrium. The results of this experiment are shown in Fig.~\ref{fig:AdiabaticRhSigmaOverview} for a cuspy (left panel) and a cored (right panel) dark matter halo. 
The snapshots shown correspond to the new equilibrium configuration reached after adiabatically lowering the (dark matter) mass $M$ from $M/M_0=1$ to a value of $\log_{10} M/M_0=0$, $-0.2$, $\dots$, $-2.0$. Snapshots for the stellar models with initial sizes of $r_\mathrm{h0} / r_\mathrm{mx0}=1$, $1/4$, $1/16$ and $1/64$ are shown in dark blue, light blue, orange, and red, respectively. The gray curves show the average velocity dispersion $\langle \sigma^2 \rangle^{1/2}$ expected for a mono-energetic stellar distribution with half-light radius $\rh$.

We note that in Fig.~\ref{fig:AdiabaticRhSigmaOverview}, all four stellar models evolve along curves of
\begin{equation}
 \rh \langle \sigma^2 \rangle^{1/2} =\text{const}~,
  \label{eq:rhsigma_invariant}
\end{equation}
indicated by a black dashed line, and all star particles remain bound throughout the adiabatic evolution. This observation implies that the bulk phase-space density is conserved during the adiabatic evolution,
\begin{equation}
 \langle Q \rangle \equiv N_\star \rh^{-3} \langle \sigma^2 \rangle^{-3/2} =\text{const}~.
\end{equation}
With no torques acting on the stellar systems studied here, single-particle angular momenta are conserved, 
\begin{equation}
 \bm{J} =  \bm{r} \times \bm{v}  = \text{const}~.
 \label{eq:angmom}
\end{equation}
For the case of mono-energetic stellar tracers embedded deeply in the density cusp or core of the underlying potential, it is straightforward to show that the product $\rh \langle \sigma^2 \rangle^{1/2}$ is a measure of the (conserved) average angular momentum
\begin{eqnarray}
\langle J^2 \rangle &=& \frac{1}{N_\star} \iint\limits_{\mathrm{all}~r,v}  | \bm{r} \times \bm{v} |^2 ~f_\star(\E)~\diff^3 r~\diff^3 v ~.
\label{eq:lum_avg_angmom}
\end{eqnarray}
where we have normalized the stellar distribution function so that $N_\star \equiv \iint f(\E) \diff^3 r~\diff^3 v$. For a mono-energetic stellar population with isotropic kinematics, embedded deeply within the power-law cusp of a Hernquist halo ($r_\mathrm{t} \ll a$ and $\E_\star \ll 1$), the integral in Eq.~\ref{eq:lum_avg_angmom} gives
\begin{equation}
\langle J^2 \rangle =  \text{const} \stackrel{\gamma=1}{=}  \frac{24}{143}  ~r_\mathrm{t}^2 ~ \langle \sigma^2 \rangle~,
\label{eq:AdiabaticModelCusp}
\end{equation}
where $r_\mathrm{t}$ is the stellar truncation radius as defined previously (Eq.~\ref{eq:HernquistStellarTruncation}), and $\langle \sigma_\star^2 \rangle$ is the average velocity dispersion,
\begin{equation}
 \label{eq:LumAveragedDispersion}
 \langle \sigma^2 \rangle = \frac{1}{N_\star}  \iint\limits_{\mathrm{all}~r,v}  v^2 ~f_\star(E)~\diff^3 r~\diff^3 v  \stackrel{\gamma=1}{=} -\frac{2}{3} \,\Phi_0\,\E ~.
\end{equation}
Similarly, for a stellar population embedded deeply in the constant-density core of a $\gamma=0$ Dehnen model, we get
\begin{equation}
\langle J^2 \rangle \stackrel{\gamma=0}{=} \frac{3}{16} ~ r_\mathrm{t}^2 ~ \langle \sigma^2 \rangle ~~~\text{where}~~~\langle \sigma^2 \rangle \stackrel{\gamma=0}{=} -\Phi_0\,\E ~.
\label{eq:AdiabaticModelCore}
\end{equation}
In the power-law limit of $r_\mathrm{t} \ll a$, we have $r_\mathrm{t} \propto r_\mathrm{h}$ (see also Fig.~\ref{fig:E_vs_rh_relation}), and the product $r_\mathrm{h} \langle \sigma^2 \rangle^{1/2} = \mathrm{const}$, consistent with the simulation results of Fig.~\ref{fig:AdiabaticRhSigmaOverview}.

\subsection{Approximate evolutionary tracks for the adiabatic case}
\label{sec:adiabatic_tracks}

A particularly intuitive way to express the adiabatic evolution of stellar components is to show the relative change in size, $\rh / r_\mathrm{h0}$, as a function of the fraction of remnant (dark matter) mass enclosed within the current half-light radius, $M_\mathrm{h} / M_\mathrm{h0}$. 
Approximating $\langle \sigma^2 \rangle \approx G M_\mathrm{h} r_\mathrm{h}^{-1}$, Eq.~\ref{eq:AdiabaticModelCusp} and \ref{eq:AdiabaticModelCore} give
\begin{equation}
 \rh / r_\mathrm{h0} \approx \left( M_\mathrm{h} / M_\mathrm{h0} \right)^{-1}~.
\end{equation}
This relation is shown, together with the corresponding simulation data, in the bottom panel of Fig.~\ref{fig:AdiabaticCollapsed}. 
Similarly for the velocity dispersion,
\begin{equation}
 \langle \sigma^2 \rangle^{1/2}/ \langle \sigma_0^2 \rangle^{1/2} \approx M_\mathrm{h} / M_\mathrm{h0} ~,
\end{equation}
and for the characteristic time $T_\mathrm{h} = 2 \pi \rh^{3/2} M_\mathrm{h}^{-1/2} G^{-1/2} $,
\begin{equation}
 T_\mathrm{h} / T_\mathrm{h0} \approx \left( M_\mathrm{h} / M_\mathrm{h0} \right)^{-2}.
\end{equation}
To summarize, for adiabatic perturbations to the underlying potential, as (dark matter) mass is lost, the stellar components expand, cool, and their characteristic dynamical times increase.

\subsection{Adiabatic evolution of the energy distribution}
\label{subsec:energy_adiabatic}

\begin{figure}[tb]
 \centering
 \includegraphics[width=8.5cm]{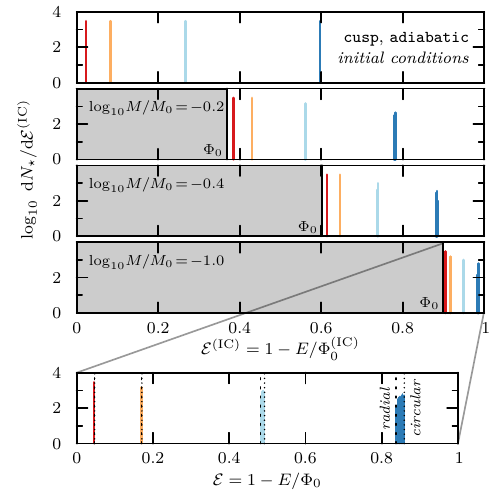} 
\caption{Equilibrium energy distributions of the four stellar profiles embedded in the cuspy halo shown in the right-hand panel of Fig.~\ref{fig:AdiabaticRhSigmaOverview}. Energies are expressed in units of the initial potential minimum, $\E^\mathrm{(IC)} = 1 - E/\Phi_0^\mathrm{(IC)} $. The top panel shows the initial conditions (IC), while the three subsequent panels correspond to snapshots where the mass $M$ sourcing the potential has been slowly lowered by -0.2, -0.4 and $-1.0,\dex$ with respect to the initial value $M_0$. The value of the evolved potential minimum $\Phi_0$ is shown in each panel, and the inaccessible region of energies below $\Phi_0$ is shaded in gray.
Bottom panel: energy distribution for the model with $\log_{10} M/M_0=-1.0$, with energies expressed in units of the evolved potential minimum $\E = 1 - E/\Phi_0$. The slight broadening of the energy distribution is caused by the difference in response of radial and circular orbits to the adiabatic perturbation. Dashed and dotted vertical lines are computed from the conservation of radial action (Eq.~\ref{Eq:RadialAction}) and angular momentum (Eq.~\ref{eq:angmom}), respectively, and accurately delimit the energy distributions measured in the $N$-body snapshots. The same systematics are observed in the case of the cored halo (not shown).}
\label{fig:AdiabaticEnergyEvolution}
\end{figure}

The four stellar populations studied in this work have mono-energetic initial conditions. In Fig.~\ref{fig:AdiabaticEnergyEvolution}, we show histograms of the energy distributions for adiabatic changes to the underlying potential, defining energies with respect to the potential minimum in the initial conditions, $\E^\mathrm{(IC)} = 1- E/\Phi_0^\mathrm{(IC)}$. With this definition, $\E^\mathrm{(IC)} = 1$ corresponds to the bound/unbound limit ($E=0$) for both the initial conditions and the evolved potential. The top panel of Fig.~\ref{fig:AdiabaticEnergyEvolution} shows the initial energy distributions of four stellar components embedded in a cuspy dark matter halo, with initial half-light radii of $r_\mathrm{h0}/r_\mathrm{mx0}=1$, $1/4$, $1/16$ and $1/64$ shown in dark blue, light blue, orange, and red, respectively. The subsequent panels show the energy distributions in the evolved potential, where the mass sourcing the potential has been adiabatically lowered to $\log_{10} M/M_0=-0.2$, $-0.4$, $-1.0$. The evolved potential minimum $\Phi_0$ is marked in each panel, and the (inaccessible) range of energies below that minimum is colored gray. The energy distributions remain narrow throughout their evolution. The energetic separation of the four mono-energetic stellar components decreases as (dark matter) mass is removed, but the stellar distributions remain energetically well separated. 

The bottom panel of Fig.~\ref{fig:AdiabaticEnergyEvolution} shows the final equilibrium energy distributions for the simulation with $\log_{10} M/M_0=-1.0$ in units of the evolved potential minimum $\E = 1 - E/\Phi_0$. Here, we note a slight broadening of the energy distribution for the more extended tracers with $r_\mathrm{h0}/r_\mathrm{mx0}= 1/16$ and $1/64$. To gain some insight into the reason for the broadening of the distribution, we consider separately the case of circular and radial orbits within the adiabatically evolving potential. For the case of adiabatically evolving spherical potentials, the invariants \citep[see e.g.][]{Weinberg1994} are the angular momentum $J$~(Eq.~\ref{eq:angmom}) and the radial action~$I$ (the latter drifts at second order with the rate of change of the underlying potential, see \citealt{Penarrubia2013} and \citealt{Burger2021} for details), 
\begin{equation}
 I = \frac{1}{\pi} \int \limits_{r_\mathrm{peri}}^{r_\mathrm{apo}}\sqrt{ -2 \Phi_0 \left[\E - \psi(r)\right]  - J^2 / r^2  } ~\diff r \approx \text{const}~,
 \label{Eq:RadialAction}
\end{equation}
where $r_\mathrm{peri}$ and $r_\mathrm{apo}$ denote the peri- and apocentre of a particle with energy $\E$ in the evolving potential. 
For circular orbits with radius $r_\mathrm{c}$, the radial action is zero, $I=0$, and angular momentum conservation dictates that $r_\mathrm{c} M(<r_\mathrm{c})=\text{const}$. The corresponding adiabatically evolved energies are indicated by vertical dotted lines in the bottom panel of Fig.~\ref{fig:AdiabaticEnergyEvolution}. For the case of radial orbits, the angular momentum is zero ($J=0$), and the approximate conservation of radial actions in slowly changing potentials requires evolved energies as indicated by the dashed vertical lines in the bottom panel of Fig.~\ref{fig:AdiabaticEnergyEvolution}. For stellar tracers that expand within a region of constant power-law slope of the underlying potential, the evolved energies of circular and radial orbits coincide, as is the case for the most deeply embedded stellar components shown in orange and red in Fig.~\ref{fig:AdiabaticEnergyEvolution}.

As an example, we explicitly show the coinciding of evolved energies for the case of stellar orbits deeply within ($\E \ll 1$) the cuspy potential. For a circular orbit of radius $r_\mathrm{c}$ within the power-law cusp, $\E = (3/2)~r_\mathrm{c}/a$ and
\begin{equation}
 J_{\gamma=1} \stackrel{\substack{\E \ll 1}}{\approx}  a \sqrt{-\Phi_0} ~ \left(2\E/3\right)^{3/2}~~~\text{(circular orbit)}~.
\end{equation}
Angular momentum conservation requires that, for adiabatic changes to the dark matter mass $M$ that sources the underlying potential, the relative change in energy $\E$ needs to scale as 
\begin{equation}
 \frac{\E}{\E_0} = \left( \frac{M}{M_0} \right )^{-1/3}~,
 \label{eq:circ_orbit_adiabatic_cusp}
\end{equation}
where the subscript zero indicates the initial state. 
Similarly, for purely radial orbits in the power-law cusp, the radial action becomes
\begin{equation}
 I_{\gamma=1} \stackrel{\substack{\E \ll 1}}{\approx}  \frac{a\sqrt{-\Phi_0}}{3\pi} \left(2\E\right)^{3/2} ~~~\text{(radial orbit)}.
\end{equation}
Adiabatic invariance of $I$ for radial orbits implies the same scaling of evolved energy $\E$ and dark matter mass $M$ as found for circular orbits using angular momentum conservation (Eq.~\ref{eq:circ_orbit_adiabatic_cusp}). Hence, the evolved energies coincide for radial and circular orbits in the power-law potential with $\gamma=1$. This is the case of self-similar adiabatic expansion.

If, however, an initially mono-energetic stellar tracer expands into a region of changing power-law slope of the underlying potential, the energies in the new equilibrium configuration differ between particles on radial and circular orbits. This causes a slight broadening of the overall energy distribution, as shown in the bottom panel of Fig.~\ref{fig:AdiabaticEnergyEvolution}. Coloured histograms show the energy distributions measured in the $N$-body snapshots, while vertical dashed and dotted lines are computed from the conservation of radial action (Eq.~\ref{Eq:RadialAction}, for radial robits) and angular momentum  (Eq.~\ref{eq:angmom}, for circular orbits), respectively. The width of the energy distribution is accurately delimited by these vertical lines.

\begin{figure*}
 \centering
\includegraphics[width=8.5cm]{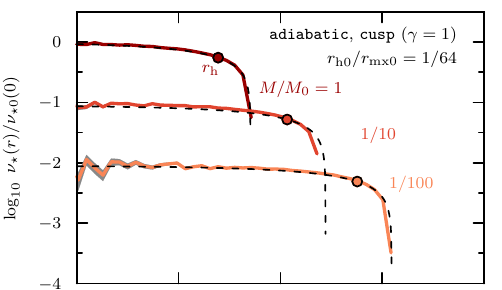}\hspace{0.9cm}\includegraphics[width=8.5cm]{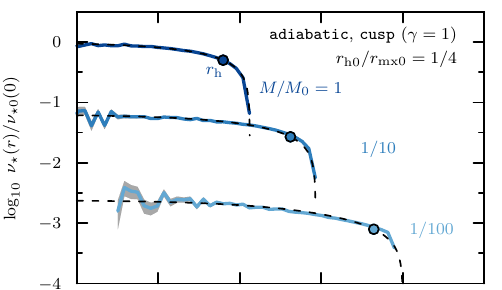} 

\includegraphics[width=8.5cm]{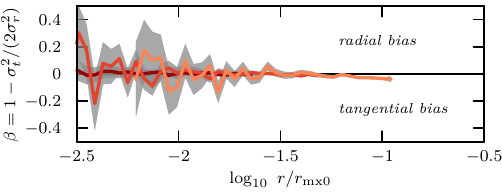}\hspace{0.9cm}\includegraphics[width=8.5cm]{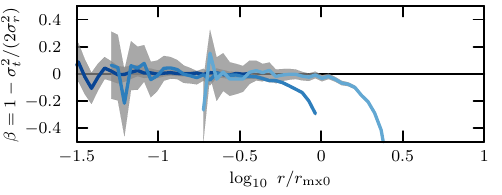} 
\caption{Number density profiles $\nu_\star(r)$ of stellar tracers with (initial) delta function energy distributions, embedded in a cuspy dark matter halo. Profiles are shown for the initial conditions ($M/M_0 = 1$), as well as for snapshots where the mass sourcing the potential as been adiabatically lowered to $M/M_0 = 1/10$ and $1/100$, respectively. The profiles are normalized by the initial central number density $\nu_{\star0}(0)$. Left panel: Stellar tracer with initial half-light radius $r_\mathrm{h0}/r_\mathrm{mx0} = 1/64$, deeply embedded within the density cusp of the surrounding halo. The density profile remains well-approximated by Eq.~\ref{eq:StellarCuspy} throughout the entire adiabatic evolution, consistent with the energy distribution remaining narrow and the kinematics remaining approximately isotropic. Right panel: Stellar tracer with $r_\mathrm{h0}/r_\mathrm{mx0} = 1/4$. As the stellar tracer expands into a region of steeper power-law slope of the density profile of the underlying halo (see bottom panel of Fig.~\ref{fig:DMdensities}), the inner regions retain isotropic kinematics, whereas at larger radii, the stellar velocity distribution becomes tangentially biased. The same systematics are observed in the case of the cored halo (not shown). }
\label{fig:AdiabaticDenisties}
\end{figure*}

\subsection{Adiabatic evolution of density and anisotropy profile}
\label{sec:adiabatic_density_profiles}
As discussed in Sec.~\ref{subsec:energy_adiabatic}, the energy distribution of a stellar component remains approximately mono-energetic for adiabatic changes to the underlying potential, as long as the component expands within a region with constant power-law slope of the underlying halo. 
In this case, the density profile of the evolved stellar distribution remains well-approximated by the functional forms of Eq.~\ref{eq:StellarCuspy} and Eq.~\ref{eq:StellarCored}, for a cuspy and a cored dark matter halo, respectively. As an example, in the left-hand panel of Fig.~\ref{fig:AdiabaticDenisties}, we show this using simulation results for a stellar tracer with initial half-light radius $r_\mathrm{h0}/r_\mathrm{mx0}=1/64$ embedded in a cuspy dark matter halo. The simulations shown here use $N=10^6$ particles.
The bottom panel of Fig.~\ref{fig:AdiabaticDenisties} shows the evolution of the anisotropy parameter $\beta(r) = 1 - \sigma^2_\mathrm{t} / (2\sigma^2_\mathrm{r})$, where by $\sigma^2_\mathrm{t}$ and $\sigma^2_\mathrm{r}$ we denote the tangential and radial components of the velocity dispersion. The stellar tracer with $r_\mathrm{h0}/r_\mathrm{mx0}=1/64$ retains near-isotropic kinematics throughout the expansion.

In contrast, the right panel of Fig.~\ref{fig:AdiabaticDenisties} shows the evolved density profiles for a more extended stellar tracer with initial $r_\mathrm{h0}/r_\mathrm{mx0}=1/4$. This tracer expands into a region with a steeper power-law slope of the underlying potential, and its energy distribution (slightly) broadens. The evolved stellar profile remains centrally isotropic but develops a tangential bias toward the outskirts. Some intuition on the origin of the tangential anisotropy can be gained from the energy distribution in the new equilibrium state. Circular orbits dominate the least-bound end of the evolved energy distribution (see Fig.~\ref{fig:AdiabaticEnergyEvolution}), whereas radial orbits delimit the most-bound end of the energy distribution, which is a consequence of the conservation of angular momentum and radial action, respectively (see Sec.~\ref{subsec:energy_adiabatic} for details). The overabundance of circular orbits at the least-bound energies manifests itself as a tangential bias in the velocity dispersion at large radii.

\subsection{Adiabatic evolution of metallicity gradients}
\label{sec:adiabatic_gradients}
Building on the results outlined in the previous subsections, we now discuss the evolution of metallicity gradients in dSph galaxies subject to adiabatic changes to the underlying potential, assuming that the total number of stars is conserved. As previously, we approximate stellar populations as initially mono-energetic distributions. We now assign a single initial metallicity $\FeH$ to each mono-energetic stellar population. As an example, we will think of the most deeply embedded population with $r_\mathrm{h0} / r_\mathrm{mx0}=1/64$ as the ``metal rich'' population (shown in red in Fig.~\ref{fig:AdiabaticRhSigmaOverview}), and similarly of the extended population with $r_\mathrm{h0} / r_\mathrm{mx0}=1$ as the ``metal poor'' one (shown in blue). For simplicity, we will assume that these initially assigned metallicities are constant in time and that the only evolution in the metallicity gradient stems from the evolution of the half-light radii of the distinct populations. For the case of a slowly decreasing dark matter mass $M/M_0$ that sources the underlying potential, the evolution of the stellar half-light radii follows curves of $M_\mathrm{h} r_\mathrm{h} \approx \text{const}$ (see Sec.~\ref{sec:adiabatic_tracks}) and is shown in Fig.~\ref{fig:AdiabaticRhSigmaOverview}. 

For two stellar populations that adiabatically expand within a region of constant power-law slope of the underlying halo (i.e., in the case of self-similar adiabatic expansion of Sec.~\ref{sec:adiabatic_density_profiles}), the metallicity gradient between the populations becomes shallower as $M/M_0$ drops, 
\begin{equation}
 \left. \frac{\Delta \FeH}{\Delta \rh}  \middle/ \frac{\Delta \FeH_0}{\Delta r_\mathrm{h0}} \right.  = \frac{\Delta r_\mathrm{h0}}{\Delta r_\mathrm{h}}  \stackrel{\rh \ll a}{\approx} M/M_0 ~,
\end{equation}
where we have used subscript zeroes to denote initial values. 

Instead, for the case of the metallicity gradient between the extended metal poor population (dark blue symbols in Fig.~\ref{fig:AdiabaticRhSigmaOverview}) and the deeply embedded metal rich one (red symbols), $|\Delta \FeH / \Delta \rh|$ will decrease even more rapidly with decreasing $M/M_0$: the more extended (metal poor) population expands in a region of steeper dropping power-law slope of the underlying dark matter halo than the deeply embedded (metal rich) one. Hence, to maintain $\rh M_\mathrm{h}=\text{const}$, the extended population expands by a larger factor than the deeply embedded one, causing the gradient between the populations to become even shallower (for a discussion of the formation of shallow metallicity gradients as a consequence of tides, see also \citealt{Ogiya2022}).

%% file: 4_impulsive.tex
\begin{figure*}[tb]
 \centering
 \includegraphics[width=18cm]{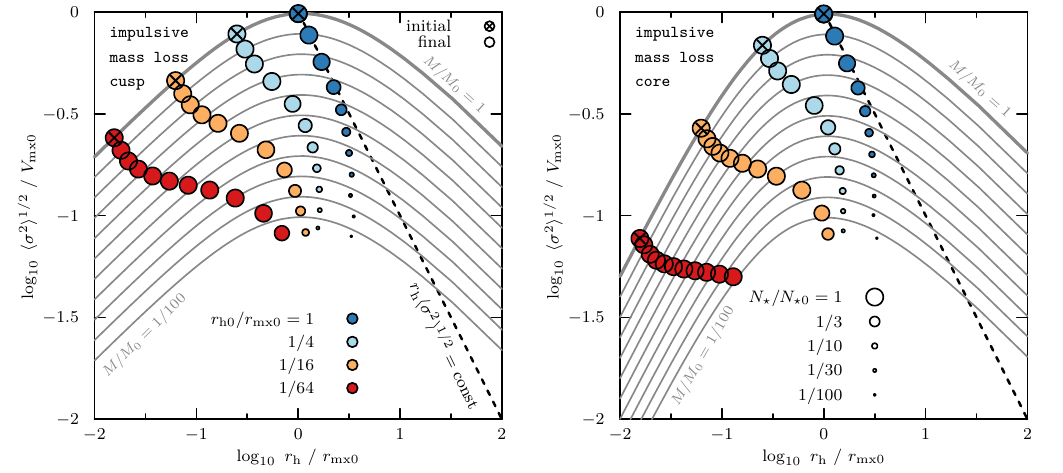} 
\caption{Like Fig.~\ref{fig:AdiabaticRhSigmaOverview} but showing the equilibrium half-light radii $\rh$ and average velocity dispersions $\langle \sigma^2 \rangle^{1/2}$ reached after an impulsive change to the underlying potential. All simulations start from the same initial conditions ($\crossedoutcircle{}$) that lie along the thick gray curve labeled $M/M_0=1$. The mass sourcing the underlying potential is then lowered, in a single step, by $0.2\,\dex$, $0.4\,\dex$, $\dots$, $2\,\dex$, and the system is allowed to return to an equilibrium state ($\emptycircle{}$). The area of the individual points corresponds to the fraction of remnant (bound) stars $N_\star/N_{\star0}$. Half-light radii and velocity dispersions are measured considering bound ($\E < 1$) stars only. For reference, the adiabatic evolution is shown as a dashed curve. In the impulsive regime, the expansion of deeply embedded stellar population is accompanied by a smaller reduction in the velocity dispersion than in the adiabatic case. The stellar populations rapidly lose stars as their half-light radii $\rh$ approach the radius of maximum circular velocity $r_\mathrm{mx}$ of the underlying dark matter halo.}
\label{fig:ImpulsiveRhSigmaOverview}
\end{figure*}

\begin{figure}[tb]
 \centering
 \includegraphics[width=8.5cm]{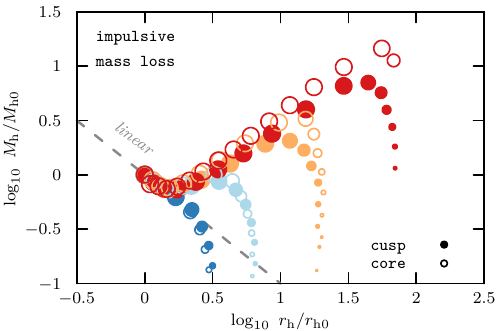} 
\caption{Like Fig.~\ref{fig:AdiabaticCollapsed} (bottom panel), but showing the relative change in remnant (dark matter) mass $M_\mathrm{h}=M(<\rh)$ (enclosed within the new equilibrium half-light radius $\rh$ reached after an impulsive change to the underlying potential) as a function of the relative change in half-light radius~$\rh / r_\mathrm{h0}$. 
In addition to the data shown in Fig.~\ref{fig:ImpulsiveRhSigmaOverview}, here we include simulations for even stronger perturbations, spanning $1 \geq M/M_0 \geq 1/1000$. The area of individual points indicates the remnant number of (bound) stars $N_\star/N_{\star0}$. In the linear regime, the final equilibrium states coincide with the adiabatic case. For the nonlinear regime, the mass enclosed within the stellar half-light radius increases as the half-light radius expands, until stars are lost.  }
\label{fig:ImpulsiveCollapsed}
\end{figure}

\section{Impulsive perturbations}
\label{sec:Impulsive}
Having studied the response of stellar systems to a slowly evolving background potential, we will now turn our attention to impulsive changes to the underlying potential. For this task, we perform numerical experiments using the same setup as in Sec.~\ref{sec:Adiabatic}, but here, we \emph{instantaneously} decrease the mass $M$ sourcing the potential. We defer the discussion of an impulsive increase of the mass, i.e. an impulsive deepening of the potential, to Appendix~\ref{sec:deepening}.

\subsection{Impulsive evolution of size and kinematics}
\label{sec:ImpulsiveStructuralParameters}
Analogously to Sec.~\ref{sec:bulk_structure_adiabatic}, we begin the discussion of the response of stellar systems to impulsive perturbations by measuring the stellar half-light radii $\rh$ and average velocity dispersions $\langle \sigma^2 \rangle^{1/2}$ in the new equilibrium configuration reached after the impulsive perturbation. We ensure that each stellar population has returned to virial equilibrium (within the limits of numerical resolution of our simulations) when we measure $\rh$ and $\langle \sigma^2 \rangle^{1/2}$. 
Fig.~\ref{fig:ImpulsiveRhSigmaOverview} shows the results of these experiments. The initial conditions are marked by crossed-out circles along the gray curve labeled $M/M_0=1$. All simulations start with these initial conditions. Open circles show the new equilibrium configurations reached after impulsively lowering the (dark matter) mass $M$ sourcing the potential by $-0.2, -0.4, \dots, -2\,\dex$, respectively, in a single step. The left-hand panel of Fig.~\ref{fig:ImpulsiveRhSigmaOverview} shows the equilibrium properties for four initially mono-energetic stellar components in a cuspy halo, while the right panel shows the equivalent simulation results for a cored halo. Gray curves show the velocity dispersion expected for a purely mono-energetic stellar component with isotropic kinematics, with the distance between two curves corresponding to a difference of $0.2\,\dex$ in the mass $M$.

For small changes to the underlying potential ($\log_{10} M/M_0 \gtrsim 10^{-0.4}$), the final equilibrium states are similar to the adiabatic case of Fig.~\ref{fig:AdiabaticRhSigmaOverview}, shown here for reference as a black dashed curve. For larger impulsive changes to the potential, the final equilibrium state differs from the adiabatic one. The properties of the final state do depend on both the strength of the perturbation (expressed through $M/M_0$) and on how deeply embedded the stellar population is within its dark matter halo. For the deeply embedded component with $r_\mathrm{h0}/r_\mathrm{mx0}=1/64$, the new equilibrium half-light radii increase with decreasing $M/M_0$, and the stellar velocity dispersions decrease, but less so than in the adiabatic case. The largest equilibrium half-light radius to which a stellar component can expand after a single-step impulsive changes to the underlying potential is limited: if a perturbation is sufficiently large so that the stellar half-light radius expands to a value of $\rh \sim \rmx$, some stars become gravitationally unbound. The loss in stars is accompanied by rapid cooling of the remnant bound stellar component, at near-constant half-light radius. The larger the initial half-light radius of the stellar component, the smaller the change of the underlying potential necessary to cause the loss of stars. To visualize the loss of stars, the area of the filled circles in Fig.~\ref{fig:AdiabaticRhSigmaOverview} is chosen to scale with the fraction $N_\star/N_{\star0}$ of remnant (bound) stars.

Similarly to the adiabatic case, the systematics of the new equilibrium states reached after an impulsive perturbation can be summarized by expressing the amplitude of the evolution of the underlying potential as a fraction of the remnant (dark matter) mass enclosed within the stellar half-light radius, $M_\mathrm{h} / M_\mathrm{h0}$.
Fig.~\ref{fig:ImpulsiveCollapsed} shows the final equilibrium $M_\mathrm{h} / M_\mathrm{h0}$ as a function of the relative change $\rh / r_\mathrm{h0}$ of the stellar half-light radius with respect to its initial value. For $\rh / r_\mathrm{mx0} \lesssim 1.5$, the final equilibrium state reached after an impulsive perturbation coincides with the adiabatic case. This regime, labeled ``linear'' in Fig.~\ref{fig:ImpulsiveCollapsed}, corresponds to small changes in the underlying potential. This is consistent with the analytical results discussed in \citet{RozierErrani2024}, showing that for the case of linear perturbations, the final equilibrium state has no memory of whether the perturbation was impulsive or adiabatic. For larger, ``nonlinear'' changes to the underlying potential, the final equilibrium half-light radius $\rh$ increases together with $M_\mathrm{h} / M_\mathrm{h0}$ (in contrast to the adiabatic case, where $M_\mathrm{h} / M_\mathrm{h0}$ decreases as $\rh/r_\mathrm{h0}$ increases). As previously, the size of the circles in Fig.~\ref{fig:ImpulsiveCollapsed} scales with the remnant number of (bound) stars $N_\star$. For impulsive perturbations that are sufficiently strong to result in the loss of stars, the final half-light radius $\rh$ (computed from the remnant bound stars) saturates at a value corresponding roughly to $\rh \sim \rmx$.

\begin{figure}[tb]
 \centering
 \includegraphics[width=8.5cm]{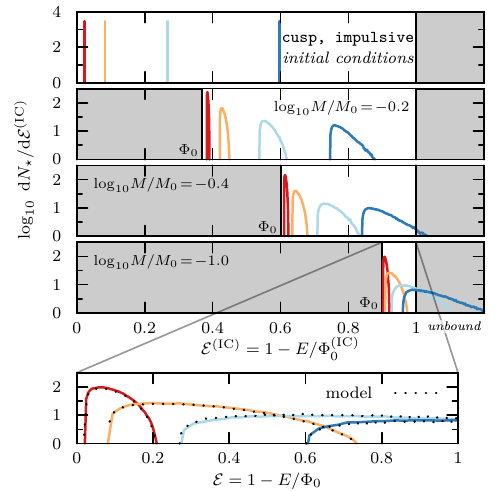} 
 
\caption{Like Fig.~\ref{fig:AdiabaticEnergyEvolution}, but showing the equilibrium energy distributions of four stellar tracers reached after an impulsive change to the underlying potential. Energies are expressed relative to the initial potential minimum. In contrast to the adiabatic case, here the stellar energy distributions widen considerably when the potential changes. These widening distributions may overlap, energetically mixing the stellar populations. Stars are lost once the stellar energy distributions extend into unbound states (gray area with $\E^\mathrm{(IC)} \ge 1$). The bottom panel shows the energy distributions measured in the simulation with $M/M_0 = 1/10$ in units of the evolved potential minimum. Dotted black curves show the energy distributions expected from the conservation of single-particle energies $\E = u^2/2 + \psi(r)$ in the evolved potential (see Eq.~\ref{Eq:EvolveddNdE}). See Fig.~\ref{fig:ImpulsiveRelativeEnergy} for a version of this plot that includes the case of impulsive deepening of the underlying potential.}
\label{fig:ImpulsiveEnergyEvolution}
\end{figure}

\subsection{Impulsive evolution of the energy distribution}
\label{sec:impulsive_energy_dist}

In contrast to the adiabatic case, stars are lost for sufficiently large impulsive changes of the underlying potential as the evolved energy distributions may extend beyond the bound/unbound limit of the underlying potential. Fig.~\ref{fig:ImpulsiveEnergyEvolution} shows the energy distributions of four stellar components with initial half-light radii of $\rh/r_\mathrm{mx0} =1$, $1/4$, $1/16$ and $1/64$ in dark blue, light blue, orange and red (as shown previously for the adiabatic case in Fig.~\ref{fig:AdiabaticEnergyEvolution}). As before, energies are expressed in units relative to the initial potential minimum, $\E^\mathrm{(IC)} = 1 - E / \Phi_0^\mathrm{(IC)}$ (see Fig.~\ref{fig:ImpulsiveRelativeEnergy} for a version of this figure with energies normalized to the evolved potential minimum). By construction, the energy distributions are delta functions in the initial conditions (top panel). As the mass sourcing the underlying potential is decreased in a single step to $M/M_0=10^{-0.2}$ (second panel from the top), $M/M_0=10^{-0.4}$ (third panel) or $M/M_0=10^{-1.0}$ (fourth panel), the energy distributions widen, and the energetic spacing between the different components decreases. For sufficiently large impulsive changes to the potential, the initially separated energy distributions may overlap in the evolved state: impulsive changes to the underlying potential can mix stellar components energetically (for an impulsive \emph{increase} of the mass sourcing the potential, see bottom panels of Fig.~\ref{fig:ImpulsiveRelativeEnergy}).

The bottom panel of Fig.~\ref{fig:ImpulsiveEnergyEvolution} shows the energy distributions for the simulation with $M/M_0=10^{-1.0}$ in units of the evolved potential minimum $\Phi_0$. Shown also are the energy distributions computed from the conservation of single-particle energies $\E = u^2/2 + \psi(r)$ in the evolved potential. Stars are treated as mass-less tracers of the underlying potential, hence this calculation is straightforward at the instant of the perturbation, where the kinetic energy can be computed from the initial potential, $u^2/2 = \left[ \Phi_0^\mathrm{(IC)} / \Phi_0 \right] ~ \left[ \E_\star^\mathrm{(IC)} - \psi^\mathrm{(IC)}(r) \right]$.  Note that while in the initial conditions the stars have a mono-energetic energy distribution, the distribution broadens in response to the impulsive perturbation: the change in potential energy of each star depends on it's position $r$ at the instant of the perturbation. For the type of impulsive perturbations studied here, the perturbed energy distribution then simply reads
\begin{equation}
 \frac{\diff N_\star}{\diff \E} =  \frac{\diff N_\star}{\diff r} ~ \left|  \frac{\diff \E}{\diff r} \right|^{-1}
 \label{Eq:EvolveddNdE}
\end{equation}
where $\diff N_\star / \diff r = 4 \pi r^2 \nu_\star(r)$ follows from the (initial, mono-energetic) stellar density profile, and $|  \diff \E / \diff r| > 0 $ is the Jacobian of the mapping between a star's position $r$ at the instant of the perturbation and it's energy $\E$ after the perturbation (for the type of perturbations studied here, the Jacobian does not vanish). 

\begin{figure*}
\centering
\includegraphics[width=8.5cm]{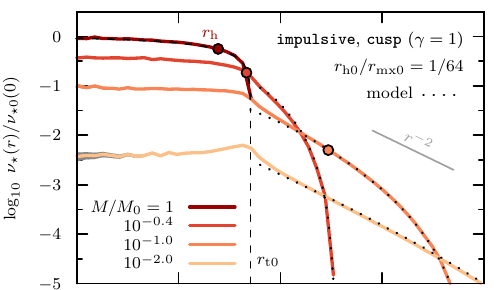}\hspace{0.7cm} \includegraphics[width=8.5cm]{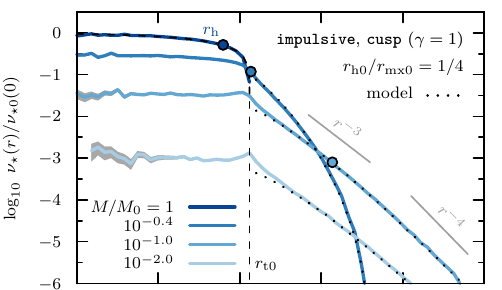}

\includegraphics[width=8.5cm]{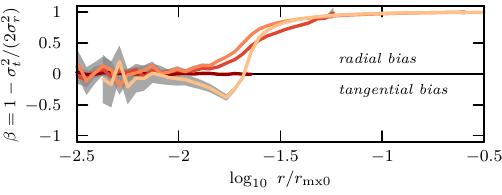}\hspace{0.7cm} \includegraphics[width=8.5cm]{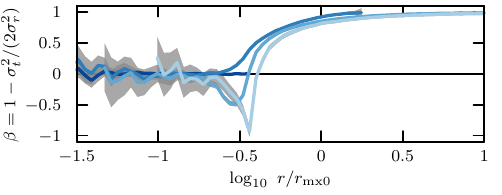}

\caption{Left-hand panels: number density profile $\nu_\star(r)$ (top) and anisotropy $\beta(r)=1 -  \sigma^2_\mathrm{t} / ( 2   \sigma^2_\mathrm{r}  )$ (bottom) of an initially mono-energetic stellar tracer with $r_\mathrm{h0}/r_\mathrm{mx0}=1/64$, embedded deeply in the density cusp of a dark matter halo. The (dark matter) mass sourcing the underlying potential is reduced impulsively, in a single step, to values of $\log_{10} M/M_0 = -0.4$, $-1.0$, $-2.0$. The density profile develops a tail toward larger radii which for $M/M_0 \lesssim 1/100$ is well-approximated by a power law. The stellar tracer retains centrally-isotropic kinematics, but is heavily radially biased in the outskirts. For the model with  $M/M_0 = 1/100$, a slight overdensity develops with tangentially-biased kinematics, located close to the initial truncation radius $r_\mathrm{t0}$ (Eq.~\ref{eq:HernquistStellarTruncation}). Dotted curves show a model (Eq.~\ref{Eq:SuperpositionGeneral}) for the outer density profile that assumes purely radial orbits for an energy distribution as in Fig.~\ref{fig:ImpulsiveEnergyEvolution} (see Sec.~\ref{sec:impulsive_density} and Appendix~\ref{appendix:mono_radial_tracer} for details). Right-hand panels: same as on the left, but for an extended stellar tracer with $r_\mathrm{h0}/r_\mathrm{mx0}=1/4$. At large radii, the log-slope of the stellar density profile approaches $\diff \ln \nu_\star(r) / \diff \ln r \rightarrow -4$. }
\label{fig:ImpulsiveDenisties}
\end{figure*}

\subsection{Impulsive evolution of the density and anisotropy profile}
\label{sec:impulsive_density}

While in the adiabatic case the shape of the stellar density profile evolves only very little, consistent with the stellar components retaining very narrow energy distributions throughout their evolution, in the impulsive case, the stellar density profile evolves considerably in response to an abrupt change of the underlying potential. 
This is shown in the left panel of Fig.~\ref{fig:ImpulsiveDenisties}, taking as an example a stellar component that initially is deeply embedded within a cuspy dark matter halo ($r_\mathrm{h0}/r_\mathrm{mx0} = 1/64$). The simulations shown here use $N=10^6$ particles, and $N=4\times10^6$ in case of the run with $\log_{10} M/M_0=-2$. As the mass sourcing the underlying potential is decreased in a single step to $\log_{10} M/M_0 = -0.4$, $-1.0$, $-2.0$, the density profile of all bound stars develops a tail toward larger radii. For stellar energy distributions that extend all the way to the bound/unbound limit ($\E=1$), the outer (bound) stellar density profile is well approximated by a gradually steepening power law. For heavily perturbed systems and large radii ($r \gg a$), we measure an outer power-law slope of $\diff \ln \nu_\star(r) / \diff \ln r   \approx -4$, consistent with simulation results of violent relaxation (see e.g. \citealt{Henon1964,vanAlbada1982} ) and the ejection of stars from globular clusters \citep{Penarrubia2017}.

Whereas the centers retain isotropic kinematics, the outskirts are radially biased (see bottom panel of Fig.~\ref{fig:ImpulsiveDenisties}). The origin of the radial bias in the outskirts is intuitively explained by noting that the impulsive perturbation studied here is spherical: single-star angular momenta are conserved and limited by the (isotropic) initial conditions. Stars that in response to the impulsive perturbation find themselves on weakly bound orbits that can reach large radii therefore have large orbital eccentricities and limited tangential velocity. 
The same systematics are observed for the case of a more extended stellar tracer with an initial half-light radius of $r_\mathrm{h0}/r_\mathrm{mx0} = 1/4$, shown in the right panels of Fig.~\ref{fig:ImpulsiveDenisties}. Note that the energy distribution of this stellar system extends into the unbound region for impulsive changes to the potential of $M/M_0=10^{-1}$ and $M/M_0=10^{-2}$ (see Fig.~\ref{fig:ImpulsiveEnergyEvolution}).

The shape of the outer density profile can be modeled to good accuracy under the assumption of purely radial orbits. The dotted curve in Fig.~\ref{fig:ImpulsiveDenisties} is calculated by integrating over a distribution of mono-energetic density profiles with purely radial kinematics (Eq.~\ref{Eq:SuperpositionGeneral}), using as weights an energy distribution $\diff N_\star / \diff \E$ as computed in Sec.~\ref{sec:impulsive_energy_dist} (Eq.~\ref{Eq:EvolveddNdE}, shown as a dotted curve in Fig.~\ref{fig:ImpulsiveEnergyEvolution}). See Appendix~\ref{appendix:mono_radial_tracer} for details. 
For stellar energy distributions that extend all the way to the bound/unbound limit, at large radii, the log-slope of the model density profile steepens to $\diff \ln \nu_\star(r) / \diff \ln r \rightarrow -4$ (see Eq.~\ref{Eq:SuperpositionOuterSlope}), consistent with the $N$-body results. As the assumption of radial orbits breaks down at radii below $r_\mathrm{t}$, we plot the model predictions for the density profiles only for radii outside of $r_\mathrm{t}$.

For both stellar tracers shown in Fig.~\ref{fig:ImpulsiveDenisties}, in the case of large impulsive changes to the underlying potential, a small overdensity emerges at the location of the initial stellar truncation radius $r_\mathrm{t0}$. This overdensity coincides with a region of tangentially-biased orbits, and consists of particles that were located in the vicinity of $r_\mathrm{t0}$, virtually at rest, at the instant of the impulsive perturbations. With (virtually) no kinetic energy at their disposal at the instant of the perturbation, these particles are confined to orbits with apocenters coinciding with $r_\mathrm{t0}$. As any remnant radial component of the velocity is exchanged for potential energy at the apocentre in the evolved potential, the remnant family of orbits sourcing the overdensity results tangentially biased (see also the phase portrait in Fig.~\ref{fig:phasemixing}, where the caustic seen at at $r_\mathrm{t0}$ corresponds to the overdensity discussed here).

\begin{figure*}
\centering
\includegraphics[width=8.5cm]{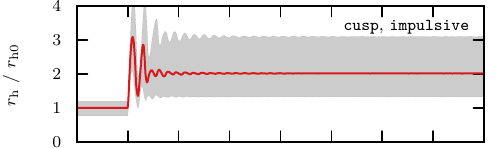}\hspace{1cm}\includegraphics[width=8.5cm]{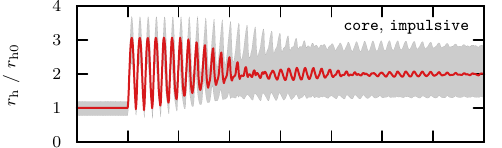}

\includegraphics[width=8.5cm]{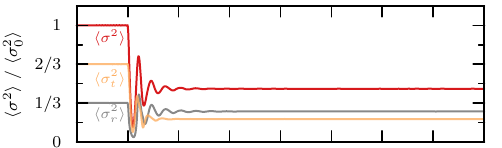}\hspace{1cm}\includegraphics[width=8.5cm]{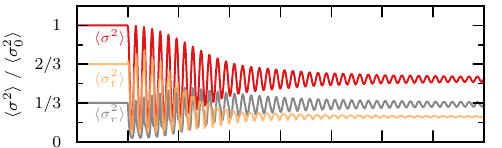}
 
\includegraphics[width=8.5cm]{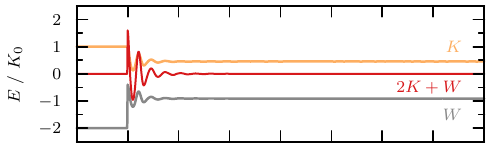}\hspace{1cm}\includegraphics[width=8.5cm]{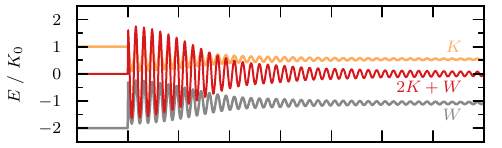}
 
\includegraphics[width=8.5cm]{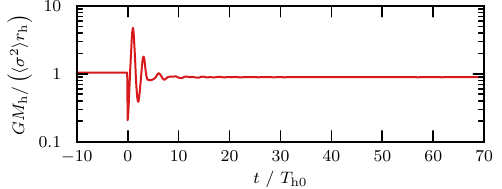}\hspace{1cm}\includegraphics[width=8.5cm]{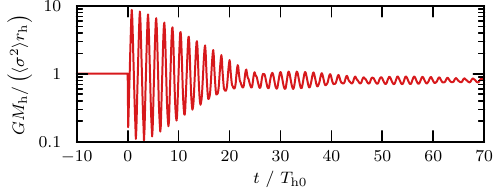}
  
\includegraphics[width=8.5cm]{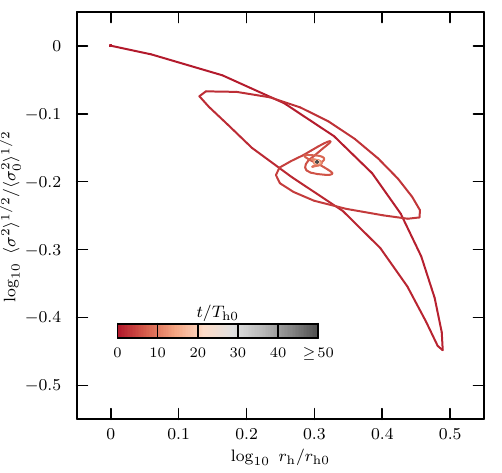}\hspace{1cm}\includegraphics[width=8.5cm]{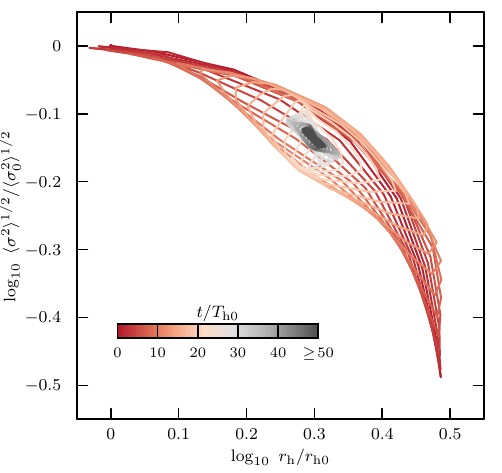}
    
\caption{Time evolution of an initially mono-energetic stellar population with initial half-light radius $r_\mathrm{h0}/r_\mathrm{mx0}=1/64$, deeply embedded in a cuspy (left) and a cored (right) dark matter halo. The dark matter mass sourcing the potential is lowered impulsively so that both stellar tracers expand by a factor of roughly two with respect to their initial half-light radius. Top row: Evolution of the half-light radius as a function of time. The $25^\mathrm{th}$ and $75^\mathrm{th}$ percentiles of the stellar radial distributions are enclosed by the gray shaded band. Times are expressed in units of the period $T_\mathrm{h0}$ of a circular orbit at the half-light radius at $t=0$. Second row: average velocity dispersion $\langle \sigma^2 \rangle$. The radial and tangential components are shown separately as gray and yellow curves, respectively. Third row: Virial quantities. total kinetic energy $T$, total potential energy $W$, and their sum $2K+W$. Fourth row: Ratio of mass $M_\mathrm{h}$ enclosed within the half-light radius and dynamically estimated mass $M_\mathrm{est}=\rh \langle \sigma^2 \rangle/G$. Bottom row: Half-light radius and velocity dispersion, normalized to their initial values. While the stellar model embedded in the cuspy halo rapidly phase mixes toward an equilibrium configuration, the stellar population embedded in the cored halo undoes several damped oscillations (in energy, size and velocity dispersion) before reaching a phase-mixed equilibrium state. For videos supplementing this figure see Appendix~\ref{Appendix:Animations}.}
\label{fig:ImpulsiveTimeEvo}
\end{figure*}

\begin{figure*}
\centering
\includegraphics[width=18cm]{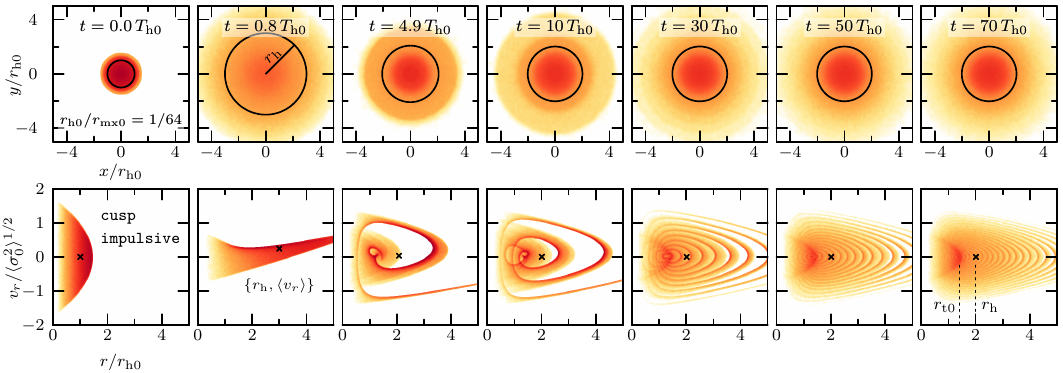}

\vspace*{-0.46cm}\hfill\textbf{(a)} Response to an impulsive perturbation of a stellar tracer deeply embedded in a cuspy halo ($r_\mathrm{h0}/r_\mathrm{mx0}=1/64$).~~\vspace*{0.6cm}

\includegraphics[width=18cm]{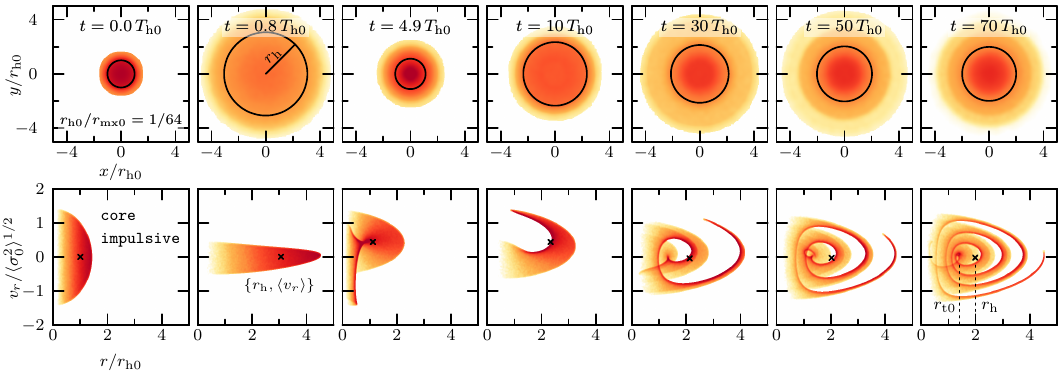}
 
\vspace*{-0.46cm}\hfill\textbf{(b)} As above, but for the case of a stellar tracer deeply embedded in a cored halo ($r_\mathrm{h0}/r_\mathrm{mx0}=1/64$).~~\vspace*{0.6cm}

 \includegraphics[width=18cm]{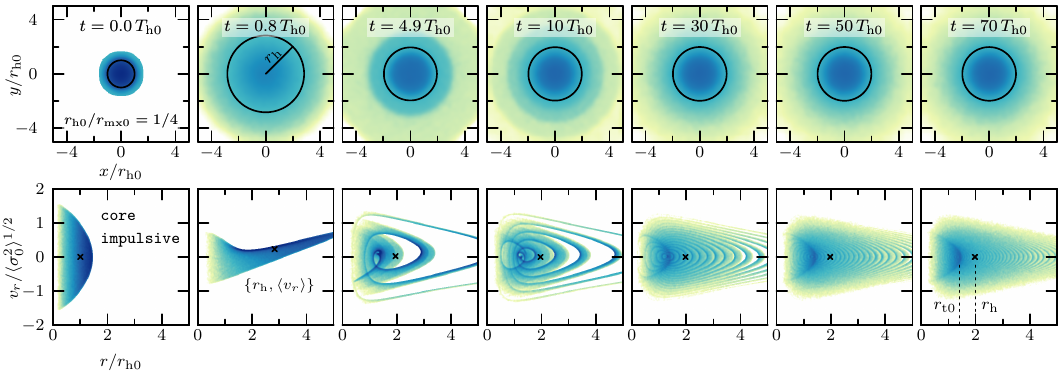}

\vspace*{-0.46cm}\hfill\textbf{(c)} As above, but for the case of an extended stellar tracer in a cored halo ($r_\mathrm{h0}/r_\mathrm{mx0}=1/4$).~~\vspace*{0.6cm}

\caption{\textbf{(a)}~Time evolution of the stellar tracer with $r_\mathrm{h0}/r_\mathrm{mx}=1/64$, deeply embedded in a cuspy halo. At $t=0$, the stellar tracer is subject to an impulsive perturbation as in Fig.~\ref{fig:ImpulsiveTimeEvo}. Top panel: (2D) surface density (colour-coded), projected onto the $\{x,y\}$-plane. A black circle with radius $R=\rh$ indicates the (3D) half-light radius. Bottom panel: radial velocity $v_r$ and (3D) radius $r$. The number of particles per bin is colour-coded. A cross indicates the location of the half-light radius $\rh$ and the median radial velocity $\langle v_r \rangle$. The stellar tracer embedded in a cuspy halo phase mixes rapidly toward an equilibrium configuration. \textbf{(b)}~Like~(a)~but for a stellar tracer deeply embedded in a cored dark matter halo. Phase mixing takes substantially longer than in the cuspy case. \textbf{(c)}~Like~(a)~but for an extended stellar tracer with initial half-light radius $r_\mathrm{h0}/r_\mathrm{mx}=1/4$, embedded in a cored dark matter halo. In contrast to (b), here, phase mixing progresses rapidly as most stars of the stellar tracer are located outside the density core and hence span a wide range of orbital periods. For videos supplementing this figure see Appendix~\ref{Appendix:Animations}.}
\label{fig:phasemixing}
\end{figure*}

\subsection{Impulsive evolution of metallicity gradients}
As discussed in Sec.~\ref{sec:adiabatic_gradients}, for the case of an adiabatic decrease of the mass sourcing the underlying potential, the metallicity gradients between stellar populations become shallower as the stellar half-light radii expand. The case of strong impulsive perturbations accompanied by the loss of (bound) stars results in qualitatively different systematics, as we briefly outline here. Analogously to Sec.~\ref{sec:adiabatic_gradients} we associate a constant metallicity to each initially mono-energetic stellar population. For the sake of argument, as previously, we refer to the most deeply embedded population as the ``metal rich'' population (see red symbols in Fig.~\ref{fig:ImpulsiveRhSigmaOverview}), and the most extended population as the ``metal poor'' population (blue symbols). 

For strong impulsive perturbations, the widening energy distribution of the extended metal poor population may extend into unbound energies, triggering the loss of (bound) stars (see Fig.~\ref{fig:ImpulsiveEnergyEvolution}). This loss of stars in turn limits the maximum half-light radius to which the remnant bound metal poor population can expand. The deeply-embedded metal rich population, however, may also expand until its energy distribution extends into unbound energies, decreasing the separation in half-light radii between the two populations. Consequently, the metallicity gradient between the remnant populations may steepen, up to the point that both populations overlap energetically and hence roughly coincide in half-light radius and velocity dispersion. In this configuration, no discernible radial dependence of the metallicity remains, and the initially distinct populations may appear as a single component with larger metallicity spread (see, e.g., the simulation with $M/M_0=1/100$ in Fig.~\ref{fig:ImpulsiveRhSigmaOverview}, where the separation between the stellar half-light radii and the velocity dispersions of all stellar populations in the final state is much narrower than in the initial conditions).

\subsection{Time evolution after impulsive perturbation}
\label{sec:impulsive_time_evo}
Having concluded the description of the final equilibrium state of a stellar population after an impulsive change to the underlying potential, we will now discuss the relaxation process through which the equilibrium state is reached. As an example, we consider a stellar tracer deeply embedded in either a cuspy or a cored potential, with an initial half-light radius of $r_\mathrm{h0}/r_\mathrm{mx0}=1/64$. We then decrease the mass sourcing the potential in such a way that the final equilibrium state in both the cuspy and the cored halo has a half-light radius of $\rh = 2r_\mathrm{h0}$. For the cuspy potential, this requires reducing the mass sourcing the potential by $M/M_0 \approx 1/25$, and for the cored case by $M/M_0 \approx 1/9$. The simulations discussed here use $N=10^6$ particles.

The top panels of Fig.~\ref{fig:ImpulsiveTimeEvo} show the time evolution of the half-light radii in the cuspy (left) and cored potential (right). 
Times are expressed in units of the (initial) period of a circular orbit at the (initial) half-light radius, $T_\mathrm{h0} = 2 \pi r_\mathrm{h0}^{3/2} (G M_\mathrm{h0})^{-1/2}$. The gray shaded bands span the $25^\mathrm{th}$ to $75^\mathrm{th}$ percentiles of the distribution of all stellar radii. The impulsive perturbation is applied at $t=0$. 

\subsubsection{Long-lived oscillatory modes} 
The stellar populations undergo a series of damped oscillations before settling into equilibrium. Crucially, it takes substantially longer for stars embedded in the cored halo to reach equilibrium than in the cuspy case: while the stellar population in the cored halo undergoes roughly ${\sim}10$ oscillations before reaching equilibrium, the stellar population in the cuspy halo reaches equilibrium in only a couple of oscillations. The bottom panels of Fig.~\ref{fig:ImpulsiveTimeEvo} show the combined time evolution of the half-light radius $\rh$ and velocity dispersion $\langle \sigma^2 \rangle^{1/2}$, with time color-coded. Also here, we observe the slower return to equilibrium of the stellar component embedded in a cored halo compared to the cuspy one. 

Some intuition into the cause underlying the slower return to equilibrium of stellar populations deeply embedded in cored subhalos compared to cuspy ones can be gained by focusing on the orbits of individual stars. Strong impulsive perturbations reset the orbital phases of all stars: after decreasing $M/M_0$ impulsively, the majority of particles will be located at the pericentre of their new orbit. The return to equilibrium of the stellar tracer particles then reduces to phase mixing these new orbits. This process is less efficient in the central regions of cored halos, where the range of orbital frequencies is much narrower than in cuspy ones. This is illustrated in the bottom panel of Fig.~\ref{fig:E_vs_rh_relation}, which shows that the crossing times for deeply embedded stellar populations in the harmonic potential of a cored halo are nearly identical (see also \citealt{Banik2021} for a discussion of phase mixing in cored halos, and \citealt{Sridhar1989} for oscillatory modes in stellar systems).

Note that in the present work, we describe the dynamical evolution of mass-less tracer particles in a background potential, and phase mixing is the sole process that allows the system of tracer particles to return to an equilibrium state. In contrast, self-gravitating systems can exhibit mode damping that is facilitated by energy exchange between particles (see e.g. \citealt{Weinberg1994_modes}, \citet{Kandrup1998}, \citealt{NelsonTremaine1999}, \citealt{Fouvry2022} and more recently \citealt{Petersen2024}, \citealt{Dattathri2025} for discussions of Landau damping in self-gravitating stellar systems).

\subsubsection{Virial equilibrium} Oscillations also appear in the evolution of the virial quantities. This is shown in the third row from the top of Fig.~\ref{fig:ImpulsiveTimeEvo}, where we plot the evolution of the total kinetic energy ($K$, yellow) and the total potential energy ($W$, gray), as well as their sum ($2K+W$, red). Virial equilibrium ($2K+W=0$) is reached much more rapidly in the cuspy halo than in the cored one. Consequently, mass modeling based on the assumption of virial equilibrium is prone to substantial systematic error for stellar populations deeply embedded in the harmonic potentials of cored halos. We will illustrate this issue by taking as an example estimators for dynamical masses that make use of the combined measurement of velocity dispersion and size to infer an enclosed dynamical mass \citep[see e.g.][]{Walker2009, Wolf2010, EPW18}.
In the fourth panel from the top of Fig.~\ref{fig:ImpulsiveTimeEvo}, we show the ratio $M_\mathrm{h}/M_\mathrm{est}$ between (dark matter) mass $M_\mathrm{h} = M(<\rh)$ enclosed within the half-light radius and a dynamical mass estimate $M_\mathrm{est} = \rh \langle \sigma^2 \rangle /G$. As the stellar population returns to virial equilibrium, the dynamically estimated mass $M_\mathrm{est}$ and the actual enclosed mass $M_\mathrm{h}$ can differ by up to an order of magnitude. 

For the case of dSph satellites of the Milky Way (with internal dynamical times of $T_\mathrm{h} = \sqrt{3\pi/(G \rho_\mathrm{h})}$ spanning roughly $50\,\mathrm{Myr}$--$0.5\,\Gyr$, see e.g. Fig.~15 in \citealt{ENIP2022}), on eccentric orbits, a tidal perturbation occurs every orbital period ($0.5\sim3\,\Gyr$ for most Milky Way satellites, see e.g. Tab.~B4 in \citealt{Battaglia2022_Gaia}). With phase mixing in the harmonic potentials of cored halos taking ${\sim}10$ internal dynamical times, this suggests that dSph satellites may spend a significant fraction of their orbits out of equilibrium if embedded in cored dark matter subhalos.

\subsubsection{Orbital anisotropy} 
The change of orbital families after an impulsive perturbation leaves an imprint on the system-averaged velocity anisotropy. The stellar populations considered in this study initially all have isotropic kinematics, $\langle \beta \rangle = 1 - \langle \sigma_\mathrm{t}^2 \rangle / (2\langle \sigma_\mathrm{r}^2\rangle) = 0$. In response to the impulsive perturbation, the populations settle in an equilibrium state with radially biased kinematics, $2\langle \sigma_\mathrm{r}^2 \rangle > \langle \sigma_\mathrm{t}^2\rangle$. We show this in the second row of panels in Fig.~\ref{fig:ImpulsiveTimeEvo}, where we plot the time evolution of the average velocity dispersion (red curve), as well as separately the evolution of the radial and tangential components (shown as gray and yellow curves, respectively).

\subsubsection{Phase space}
The phase mixing after an impulsive perturbation leaves clear signatures in phase space (see, e.g. \citealt{Burger2019} in the context of dark matter core formation, and \citealt{Laporte2019, Banik2022, Banik2023} for perturbations to the Galactic disc). For selected simulation snapshots, in Fig.~\ref{fig:phasemixing} we show the surface brightness $\Sigma_\star(x,y)$ (top panels), as well as a projection of phase space using spherical radius $r$ and radial velocity $v_r = \bm{v} \cdot \bm{r}/r$ as coordinates (bottom panels). Fig.~\ref{fig:phasemixing}a and Fig.~\ref{fig:phasemixing}b show a stellar population with $r_\mathrm{h0}/r_\mathrm{mx0}=1/64$, deeply embedded in a cuspy and a cored halo, respectively. These are the same simulations as shown in Fig.~\ref{fig:ImpulsiveTimeEvo} (left panels show the cuspy model, right panels the cored one). 
The stellar populations return to equilibrium after an impulsive perturbation through inward and outward moving shells of stars. Phase mixing washes out the coherent motions visible in phase space during the first dynamical times after the perturbation. For the case of a stellar population deeply embedded in a cored halo (Figure~\ref{fig:phasemixing}b), the phase space retains structure for many more dynamical times than in the case of the cuspy halo (Figure~\ref{fig:phasemixing}a). Note the caustic coinciding with the initial stellar truncation radius $r_\mathrm{t0}$ (separating families of orbits with apocentres above or below $r_\mathrm{t0}$) which gives rise to the slight overdensity discussed in Sec.~\ref{sec:impulsive_density}. Finally, Fig.~\ref{fig:phasemixing}c shows the return to equilibrium for a more extended stellar tracer embedded in a cored halo ($r_\mathrm{h0}/r_\mathrm{mx0}=1/4$). Here, phase mixing progresses rapidly, as outside the harmonic potential of the density core, the range of orbital frequencies is much wider than inside the core.

\subsection{Response to linear perturbations}

Finally, to make a connection with the results discussed in \citet{RozierErrani2024}, we turn our attention to the response of a stellar system to impulsive perturbations in the linear regime. As shown in Sec.~\ref{sec:bulk_structure_adiabatic}, stellar systems subject to adiabatic changes of the underlying potential evolve along curves of $\rh \langle \sigma^2 \rangle^{1/2}=\text{const}$ toward an equilibrium state. Crucially, the final equilibrium state does not depend on how it was reached but only on the total change of the underlying potential. Whether this change occurred all at once or in several discrete steps does not affect the remnant state, as long as each perturbation occurred on a time scale sufficiently long to be perceived as adiabatic. As a consequence, there is a single evolutionary track for the adiabatic case.

\begin{figure}[tb]
 \centering

  \includegraphics[width=8.5cm]{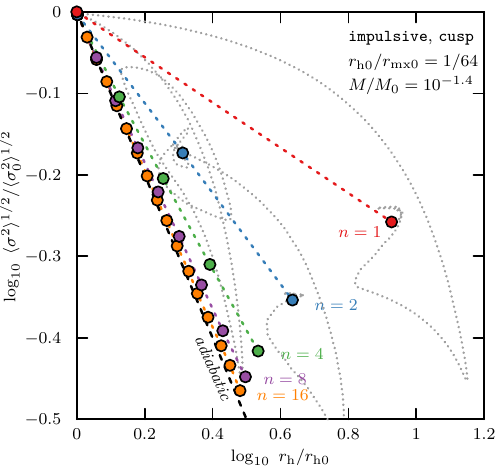} 

   \caption{Equilibrium configurations of half-light radius $\rh$ and average velocity dispersion $\langle \sigma^2 \rangle^{1/2}$ reached after an impulsive change to the underlying potential. The mass sourcing the potential is decreased to a value of $M/M_0 = 10^{-1.4}$ in $n=1$, $2$, $4$, $8$ and $16$ discrete steps (shown in red, blue, green, purple and orange, respectively). The smaller the relative change in potential between two subsequent steps (i.e., the more linear are perturbation and response), the closer is the overall evolution is to the adiabatic case. As an example, for the cases of $n=1$ and $n=2$, the full time evolution toward the equilibrium states is depicted using dotted curves.   }
   \label{fig:ImpulsiveSteps}
\end{figure}

Clearly this does not hold for impulsive perturbations. In Fig.~\ref{fig:ImpulsiveSteps}, we show the results of an experiment in which we impulsively decrease the mass sourcing the underlying potential to a final value of $M/M_0=10^{-1.4}$. We decrease the mass in $n=1$, $2$, $\dots$, $16$ logarithmically spaced steps, and take note of the intermediate equilibrium half-light radii $\rh$ and velocity dispersions $\langle \sigma^2 \rangle^{1/2}$ of an embedded stellar component. The final state for a single impulsive step ($n=1$) is shown in red, whereas the intermediate and final equilibrium states for $n=2$, $4$, $8$ and $16$ are shown in blue, green, purple, and orange. A dashed gray curve shows the full time evolution toward the final and the intermediate states for the cases with $n=1$ and $n=2$. 

We observe that as the number of steps $n$ increases, the final and intermediate states move closer and closer to the adiabatic solution. This behavior is consistent with the analytical results discussed in \citet{RozierErrani2024}: for stellar systems that are only slightly out of equilibrium, with both the perturbation and the response in the linear regime and non-resonant, when neglecting second-order terms, \citet{RozierErrani2024} found that the system's global properties in the final equilibrium state are independent of the detailed time evolution of the perturbation. In other words, in the linear, non-resonant regime, the global properties of the final equilibrium state reached by a stellar system in response to a perturbation are independent of whether the perturbation is perceived as adiabatic or impulsive. In the numerical experiments presented in Fig.~\ref{fig:ImpulsiveSteps}, for small consecutive impulsive changes to the underlying potential, both the perturbation to the stellar component as well as its response are approximately linear, and we recover an evolution that coincides with the adiabatic case.

%% file: 5_summary.tex
\section{Summary}
\label{sec:Summary}
The equilibrium state reached by a stellar system in response to a perturbation depends on the time scale on which the perturbation is applied. In this work, we use tracer particle $N$-body integrations to study the response of dwarf spheroidal galaxies embedded in dark matter halos to adiabatic and impulsive changes to the underlying (dark matter) potential. We assume an initial separation of different stellar populations in chemo-energetic space.
Our main results are summarized below. 

\begin{itemize}[leftmargin=*, itemsep=-2pt]
 \item[i)] We develop a family of mono-energetic density profiles for dwarf spheroidal galaxies with isotropic kinematics that are embedded in cuspy or cored dark matter halos. The density profiles and the velocity dispersion profiles are centrally flat and sharply truncated beyond some radius $r_\mathrm{t}$. Any spherical distribution of stars with isotropic kinematics may be constructed by superposition of these mono-energetic profiles. 
 \item[ii)] For perturbations that are perceived as adiabatic by the stellar component, half-light radii $\rh$ and average velocity dispersions $\langle \sigma^2 \rangle^{1/2}$ evolve along curves of $\rh \langle \sigma^2 \rangle^{1/2} = \text{const}$. When slowly decreasing the (dark matter) mass sourcing the potential, half-light radii expand, velocity dispersions drop, and metallicity gradients flatten.
 \item[iii)] During the adiabatic evolution, each mono-energetic stellar tracer retains centrally isotropic kinematics. Stellar tracers that expand into a region of steepening power-law slope of the underlying (dark matter) profile develop tangentially biased kinematics in their outskirts, and their binding energy distributions slightly widen.
 \item[iv)] For the case of power-law potentials, stellar binding energy distributions that were initially separated will remain separated during the adiabatic evolution: stellar populations do not mix energetically for adiabatic changes to the underlying potential.
 \item[v)] In contrast, for impulsive perturbations, stellar energy distributions widen substantially and may overlap, energetically mixing initially separate populations to the extent that initially chemo-dynamically distinct populations may masquerade as a single population with larger metallicity and energy spread.
 \item[vi)] The widening energy distributions may extend into unbound energies, resulting in the loss of stars. This limits the maximum size to which a stellar population may expand in response to impulsive changes of the underlying population. 
 \item[vii)] Stellar populations subject to strong impulsive perturbations develop power-law-like density tails with radially biased kinematics. 
 \item[viii)] Stellar populations that are deeply embedded in the constant-density core of a dark matter halo phase mix inefficiently in response to an impulsive perturbation, giving rise to a series of long-lived oscillations that may persist for tens of dynamical times. This slow return to equilibrium adds substantial uncertainty to dynamical masses estimated from Jeans modeling or the virial theorem. For the case of cuspy halos, where stellar populations span a wide range of orbital frequencies, phase mixing progresses rapidly, and a new equilibrium state is reached after only a couple of dynamical times.
 \item[ix)] Consistent with previous work, we show that as impulsive perturbations decrease in amplitude (i.e., as they become more linear), the equilibrium state reached by the stellar population approaches that of the adiabatic case. 
\end{itemize}
Our models may be applied to aid the interpretation of the response of stellar systems to a wide range of perturbations, such as, for example, tidal mass loss, impulsive and adiabatic outflows, the contraction of an underlying halo due to accretion, and the evolution of an underlying potential in core-collapsing halos for models of self-interacting dark matter.

%% file: Appendix.tex
\section{Impulsive deepening of the underlying potential}
\label{sec:deepening}
\label{appendix:MassIncrease}
In this appendix, we discuss the response of spherical stellar populations to an impulsive deepening of the underlying potential. We consider stellar populations that in the initial conditions are mono-energetic and have isotropic kinematics, with initial half-light radii of $r_\mathrm{h0} / r_\mathrm{mx0} = 1$, $1/4$, $1/16$ and $1/64$. The numerical setup is identical to that used in Sec.~\ref{sec:Impulsive}. All simulations start with the same initial condition. We then impulsively increase the mass $M$ which sources the potential, in a single step, by $+0.2\,\dex$, $+0.4\,\dex$, $\dots$, $+2.0\,\dex$, respectively. As an example, here we discuss the evolution of stellar populations in cuspy dark matter halos, but the same systematics hold also for the case of stellar tracers embedded in cored halos.

\begin{figure}[tb]
 \centering
  \includegraphics[width=8.5cm]{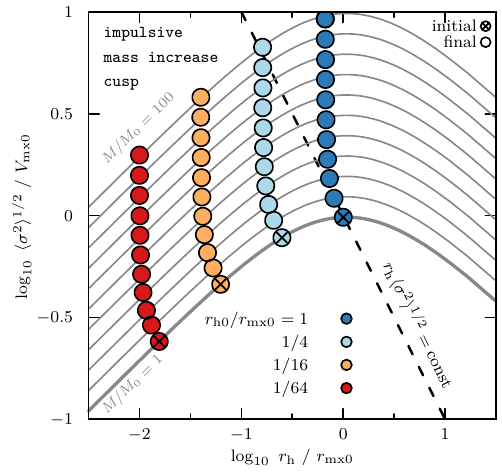}
  \caption{Like Fig.~\ref{fig:ImpulsiveRhSigmaOverview}, but showing the equilibrium half-light radius $\rh$ and velocity dispersion $\langle \sigma^2 \rangle^{1/2}$ reached after an impulsive deepening of the underlying potential. The (dark matter) mass sourcing the underlying potential is increased impulsively, in a single step, by $+0.2\,\dex$, $+0.4\,\dex$, $\dots$, $+2\,\dex$. In response to the impulsive increase in mass, the half-light radii contract, and the velocity dispersion increases. The same systematics are observed for stellar tracers embedded in a cored dark matter halo (not shown). }
  \label{fig:sigma_rh_mass_increase}
\end{figure}

\begin{figure}[tb]
 \centering
\includegraphics[width=8.5cm]{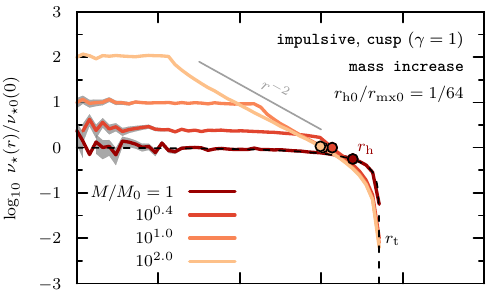}

\includegraphics[width=8.5cm]{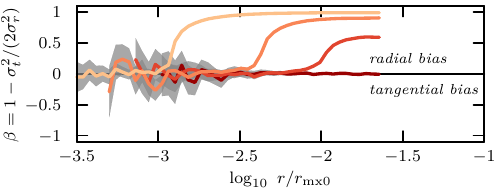}
\caption{Like Fig.~\ref{fig:ImpulsiveDenisties}, but for impulsive deepening of the underlying potential, showing the density profile (top panel) and anisotropy (bottom panel) of an initially mono-energetic stellar tracer with $r_\mathrm{h0}/r_\mathrm{mx0}=1/64$, embedded deeply in the density cusp of a dark matter halo. The stellar tracer retains centrally-isotropic kinematics, but is heavily radially biased in the outskirts. The same systematics are observed for stellar tracers embedded in a cored dark matter halo (not shown). }
\label{fig:rho_beta_mass_increase}
\end{figure}

Fig.~\ref{fig:sigma_rh_mass_increase} shows the equilibrium half-light radii $\rh$ and average velocity dispersions $\langle \sigma^2 \rangle^{1/2}$ of the four mono-energetic stellar populations in the initial conditions ($M/M_0=1$, crossed circles), as well as after an impulsive perturbation ($M/M_0 = 10^{0.2},\dots,10^{2}$, open circles). For all four populations, the half-light radii decrease in response to the perturbation, and the velocity dispersion increases. For the regime of linear (i.e., small) perturbations, the evolution follows curves of $\rh \langle \sigma^2 \rangle^{1/2}=\text{const}$. Note that the contraction of half-light radii in response to the impulsive deepening of the underlying potential is limited: for strong perturbations, where a particle's kinetic energy becomes negligible with respect to the potential energy in the deepened potential, the location of a particle at the instant of the perturbation sets its new apocentre. In this limit, further deepening of the potential does not alter the stellar half-light radii, but instead only results in an increase in the average velocity dispersion.

In response to an impulsive increase of $M/M_0$, stellar populations contract and develop a power-law density tail toward the initial stellar truncation radius $r_\mathrm{t}$, see Fig.~\ref{fig:rho_beta_mass_increase}. The evolved equilibrium density profiles are centrally flat with isotropic kinematics, and have radially biased kinematics in their outskirts. 

\begin{figure}[tb]
 \centering
  \includegraphics[width=8.5cm]{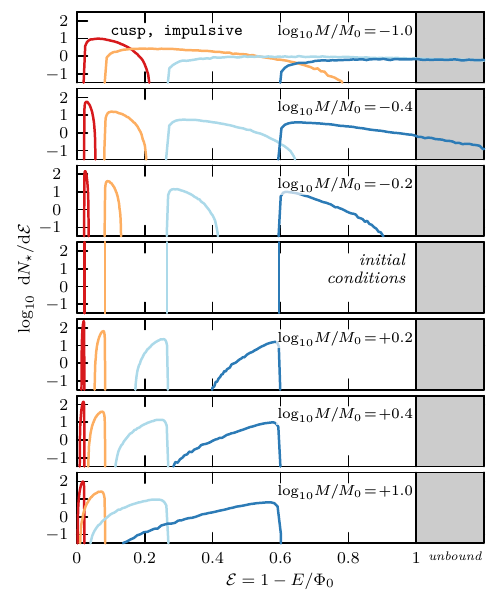}
  \caption{Like Fig.~\ref{fig:ImpulsiveEnergyEvolution}, but with energies expressed with respect to the evolved potential minimum. The evolved energy distributions are shown for impulsive reductions of the (dark matter) mass that sources the potential (top three panels), the initial conditions (central panel), and for impulsive deepening of the underlying potential (bottom three panels; see Appendix~\ref{appendix:MassIncrease} for a discussion of the equilibrium structural parameters for this scenario). For both the cases of impulsive (dark matter) mass increase and decrease, the stellar binding energy distributions widen with respect to the initial conditions, and may overlap, mixing the initially energetically separated populations. }
  \label{fig:ImpulsiveRelativeEnergy}
\end{figure}

As was the case for an impulsive reduction of the mass sourcing the potential (Sect.~\ref{sec:Impulsive}), also here, for an impulsive increase of $M/M_0$, the distribution of gravitational binding energies widens in response to the perturbation; see Fig.~\ref{fig:ImpulsiveRelativeEnergy}. Energy distributions are shown for the initial conditions (central panel), as well as for the cases of impulsive mass decrease (top three panels) and impulsive mass increase (bottom three panels).

\section{Mono-energetic tracer with purely radial orbits}
\label{appendix:mono_radial_tracer}
\begin{figure}[tb]
 \centering
  \includegraphics[width=8.5cm]{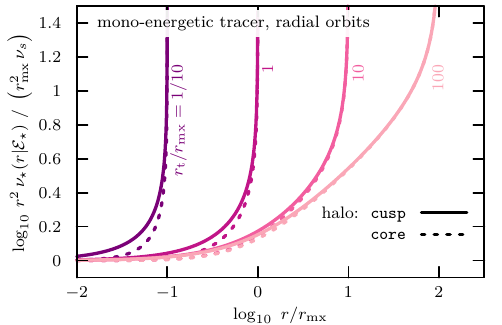}
  \caption{Number density profiles $r^2 \nu_\star(r|\E_\star)$ for mono-energetic stellar tracers with purely radial orbits (Eq.~\ref{eq:mono_radial_density}), embedded in a cuspy (solid curve, Eq.~\ref{eq:Hernquist_pot}) and a cored (dashed curve, Eq.~\ref{eq:coredPsi}) dark matter halo. Profiles are shown for four different values of the stellar truncation radius $r_\mathrm{t}/r_\mathrm{mx}=1/10$, $1$, $10$, $100$. The profiles asymptotically behave like $\nu_\star \propto r^{-2}$ for $r\rightarrow 0$, and have a caustic at the radius $r_\mathrm{t}$ caused by the pileup of stars at the apocentre $r_\mathrm{t}$.  }
  \label{fig:radial_tracer_density}
\end{figure}

Following the same steps as in Sec.~\ref{sec:NumericalSetup}, we now compute the density profile of a spherical mono-energetic stellar tracer whose stars move on purely radial orbits. Such a stellar tracer has a distribution function of the form
\begin{equation}
 f_\star(\E,J) \propto \delta(\E - \E_\star) ~ \delta(J)~,
\end{equation}
where by $\E$ we denote the energy (Eq.~\ref{eq:EnergyDef}) and by $J$ we denote the magnitude of the angular momentum (Eq.~\ref{eq:angmom}). The distribution function above is identical to zero, unless a star has angular momentum zero and energy $\E_\star$. By integrating the distribution function over all velocities, we find the density profile given by Eq.~\ref{eq:mono_radial_density}. One may reach the same result by considering that the number of stars $\diff N_\star$ located at a radius $r$ is proportional to the time each star spends at that radius along its orbit, $\diff t = \diff r / v(r)$, where $v(r)$ is the star's velocity.  
Hence, $\diff N_\star = 4 \pi r^2 \nu_\star(r|\E_\star) \diff r  \propto   \diff r / v(r)$, and the stellar number density profile reads
\begin{equation}
  \nu_\star(r|\E_\star)  =  
  \begin{dcases}
    \nu_\mathrm{s}~(a/r)^2  \big/ \sqrt{ \E_\star - \psi(r)}  & \text{if } \psi(r) < \E_\star \\
     0 ~ & \text{otherwise,}
  \end{dcases}
  \label{eq:mono_radial_density}
\end{equation}
where by $\psi(r)$ we denote the underlying gravitational potential (Eq.~\ref{eq:PotentialDef}), $a$ is a scale radius (chosen here to correspond to the scale radius of the underlying dark matter halo), and $\nu_\mathrm{s}$ is a stellar scale density.
This mono-energetic density profile is cuspy with $\diff \ln \nu_\star(r|\E_\star) / \diff \ln r \rightarrow -2$ for $r\rightarrow 0$, and has a caustic at the radius $r_\mathrm{t}$ where $\psi(r_\mathrm{t}) = \E_\star$. This radius is identical to the truncation radius discussed in Sec.~\ref{sec:NumericalSetup} (Eq.~\ref{eq:HernquistStellarTruncation}, \ref{eq:DehnenStellarTruncation}) beyond which the stellar density is zero. The caustic is caused by the (radial) velocity of stars dropping to zero at their apocentre $r_\mathrm{t}$. Fig.~\ref{fig:radial_tracer_density} shows four example stellar profiles following Eq.~\ref{eq:mono_radial_density} with $r_\mathrm{t}/r_\mathrm{mx} = 1/10$, $1$, $10$, $100$, where by $r_\mathrm{mx}$ we denote the radius of peak circular velocity of the underlying dark matter halo. Solid and dashed curves correspond to stellar profiles embedded in a cuspy (Eq.~\ref{eq:Hernquist_pot}) and a cored (Eq.~\ref{eq:coredPsi}) dark matter halo, respectively.

\begin{figure}[b]
 \fbox{\includegraphics[width=8.5cm]{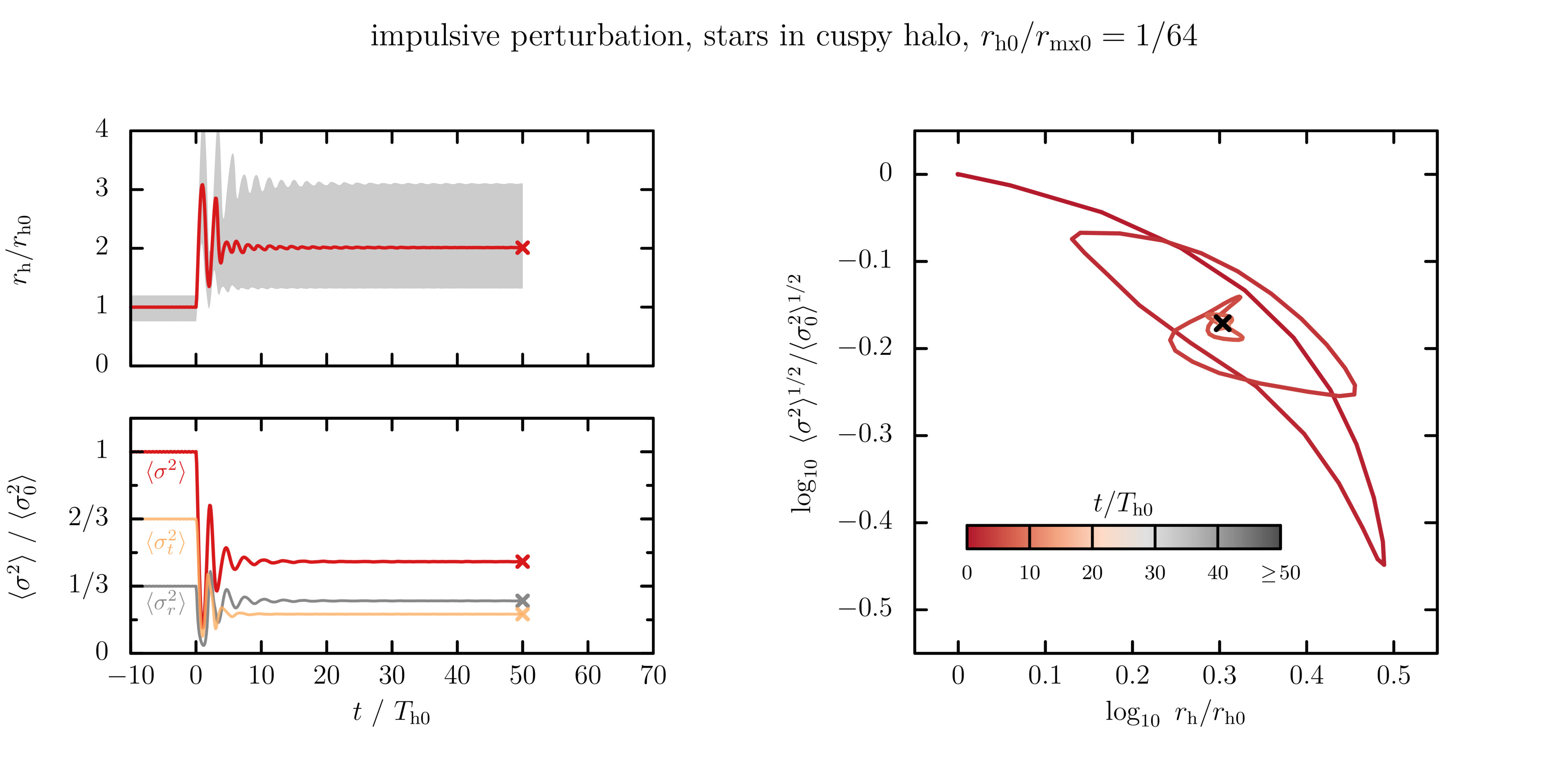}}
 \caption{Animated figure supplementing the left-hand panel of Fig.~\ref{fig:ImpulsiveTimeEvo}. The video shows the time evolution of the half-light radius and the velocity dispersion of a stellar tracer population in response to an impulsive perturbation in a cuspy dark matter halo. File size $0.9\,\mathrm{MB}$, dimensions $3840\times1920\,\mathrm{px}^2$, duration $33\,\mathrm{s}$. The animated figure is available in the \href{https://arxiv.org/src/2502.19475v2/anc}{arXiv ancillary files}.}
 \label{fig:ani_13left}
\end{figure}

\begin{figure}[b]
 \fbox{\includegraphics[width=8.5cm]{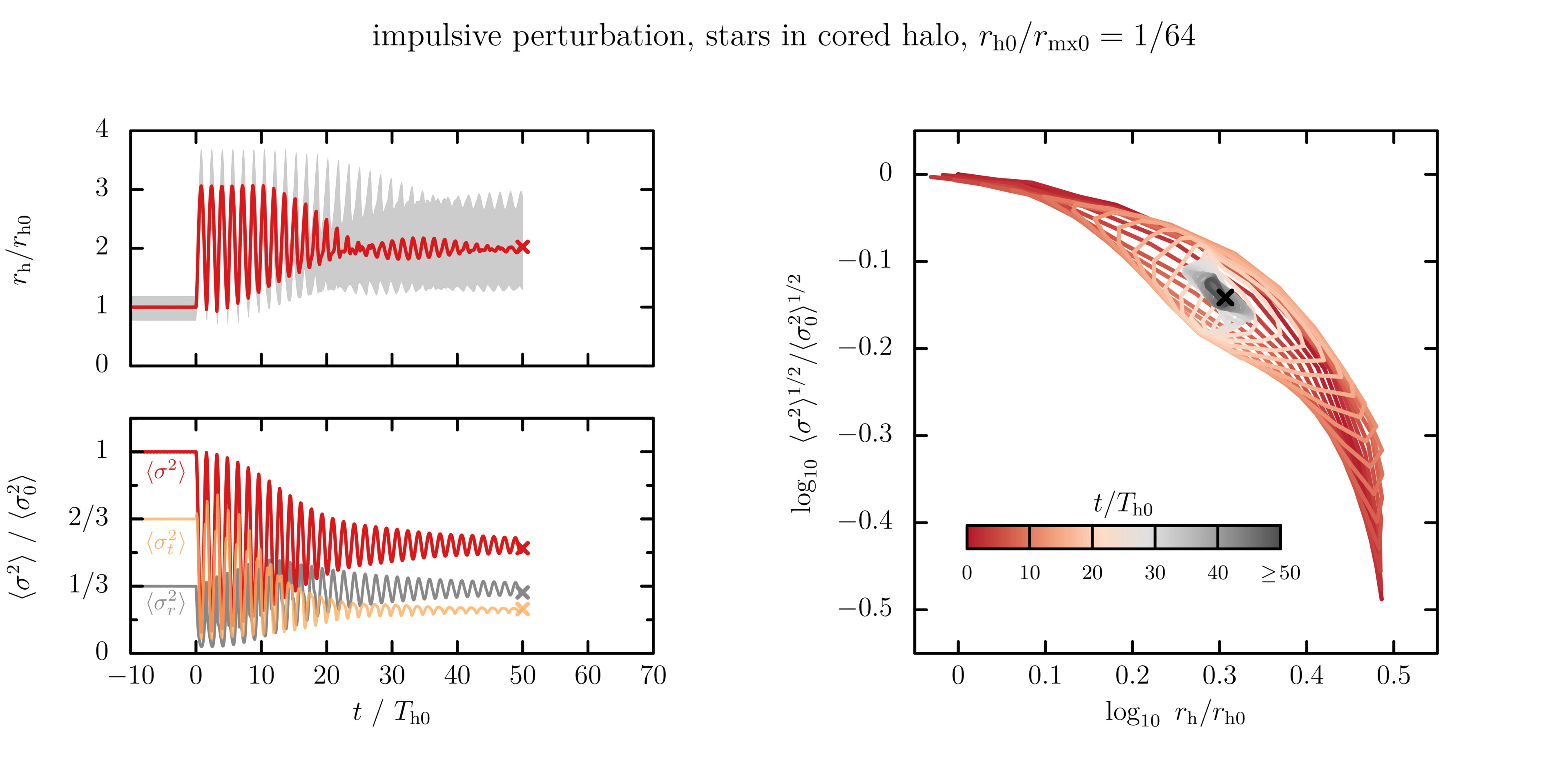}}
 \caption{Animated figure supplementing the right-hand panel of Fig.~\ref{fig:ImpulsiveTimeEvo}. The video shows the time evolution of the half-light radius and the velocity dispersion of a stellar tracer population in response to an impulsive perturbation in a cored dark matter halo. File size $1.3\,\mathrm{MB}$, dimensions $3840\times1920\,\mathrm{px}^2$, duration $33\,\mathrm{s}$. The animated figure is available in the \href{https://arxiv.org/src/2502.19475v2/anc}{arXiv ancillary files}.}
 \label{fig:ani_13right}
\end{figure}

We compute the scale density $\nu_s$ by requiring that $N_\star = 4 \pi \int_0^{r_\mathrm{t}} r^2 \nu_\star(r|\E_\star) \diff r$. For a stellar tracer of energy $\E_\star$ embedded in a cuspy ($\gamma=1$) Hernquist halo with scale radius $a=r_\mathrm{mx}$, we find
\begin{equation}
 \nu_s  \stackrel{\substack{\gamma=1}}{=}  \frac{N_\star}{4\pi a^3} ~~ \frac{(1-\E_\star)^{3/2}}{\sqrt{\E_\star-\E_\star^2} + \arctan \left( \sqrt{\frac{\E_\star}{1-\E_\star}}\right)  }  ~.
\end{equation}
For the case of a stellar tracer embedded deeply in the power-law cusp, $\E_\star \ll 1$, the equation above reduces to $\nu_s \approx N_\star  / (4\pi a^3 \sqrt{\E_\star})$.

Similarly, for a stellar tracer in a cored ($\gamma=0$) Dehnen halo with scale radius $a=r_\mathrm{mx}/2$, we find
\begin{equation}
 \nu_s  \stackrel{\substack{\gamma=0}}{=}  \frac{N_\star}{4\pi a^3} ~~ \frac{ (1-\E_\star)^{3/2} }{ \frac{\pi}{2}  + \sqrt{\E_\star - \E_\star^2}  + \arctan \left( \sqrt{\frac{\E_\star}{1-\E_\star}}\right) }  ,
\end{equation}
which, for the case of a stellar component embedded deeply in the constant-density core, $\E_\star \ll 1$, reduces to $\nu_s \approx N_\star / \left(2\pi^2 a^3 \right)$.

In Sec.~\ref{sec:impulsive_density}, we use a superposition of stellar profiles following Eq.~\ref{eq:mono_radial_density} to model the outer density profiles of stellar systems that were affected by impulsive perturbations. The superposition is computed by weighting the mono-energetic profiles $\nu_\star(r)$ by differential energy distribution $\diff N_\star / \diff \E$ of Fig.~\ref{fig:ImpulsiveEnergyEvolution}, integrating from the most-bound state $\E_\star=0$ to the bound/unbound limit $\E_\star = 1$,
\begin{equation}
   \nu_\star(r) = \int_0^1 \frac{\diff N_\star}{\diff \E_\star} ~ \nu_\star(r|\E_\star) ~ \diff \E_\star~,
   \label{Eq:SuperpositionGeneral}
\end{equation}
where each mono-energetic $\nu_\star(r|\E_\star)$ is normalized here so that $1 = 4 \pi \int_0^\infty r^2 \nu_\star(r|\E_\star) \diff r$.

For stellar energy distributions that extend all the way to the bound/unbound limit, at large radii, the log-slope of this model density profile steepens to a value of $\diff \ln \nu_\star(r) / \diff \ln r \rightarrow -4$. This can be shown by noting that, for $\E_\star \sim 1$, the density profile of a mono-energetic stellar tracer scales as $\nu_\star(r|\E_\star) \propto r^{-2} (1-\E_\star)^{3/2} / \sqrt{\E_\star - \psi(r)}$. At each radius $r \gg a$ within the density tail, only stars on orbits with energy $\E_\star \geq \psi(r)$ contribute to the local density. For a stellar energy distribution  $\diff N_\star / \diff \E_\star \approx \text{const}$ that is flat at the bound/unbound limit $\E_\star = 1$, the superposition of mono-energetic stellar density profiles at $r \gg a$ reads
 \begin{equation}
  \nu_\star(r) \!\stackrel{r \gg a}{\propto}\! \frac{1}{r^2}\!\!\int\limits_{\psi(r)}^{1}\!\!\frac{(1\!-\!\E_\star)^{3/2}}{\sqrt{ \E_\star \!-\! \psi(r)}} \, \diff \E_\star \!\propto\! \frac{[  \psi(r) \!-\! 1 ]^2}{r^2} \!\propto\! \frac{1}{r^4}. 
  \label{Eq:SuperpositionOuterSlope}
 \end{equation}
Dotted curves in Fig.~\ref{fig:ImpulsiveDenisties} (Sec.~\ref{sec:impulsive_density}) show stellar density profiles that are modeled as a superposition of mono-energetic densities (Eq.~\ref{eq:mono_radial_density}) weighted by the respective stellar energy distribution of Eq.~\ref{Eq:EvolveddNdE} (shown as a dotted curve in Fig.~\ref{fig:ImpulsiveEnergyEvolution}). For stellar tracers whose energy distributions extend all the way to the bound/unbound limit, the log-slope $\diff \ln \nu_\star(r) / \diff \ln r$ approaches $-4$ for $r \gg a$, consistent with the analytical estimate of Eq.~\ref{Eq:SuperpositionOuterSlope}.

\begin{figure}[b]
\fbox{\includegraphics[width=8.5cm]{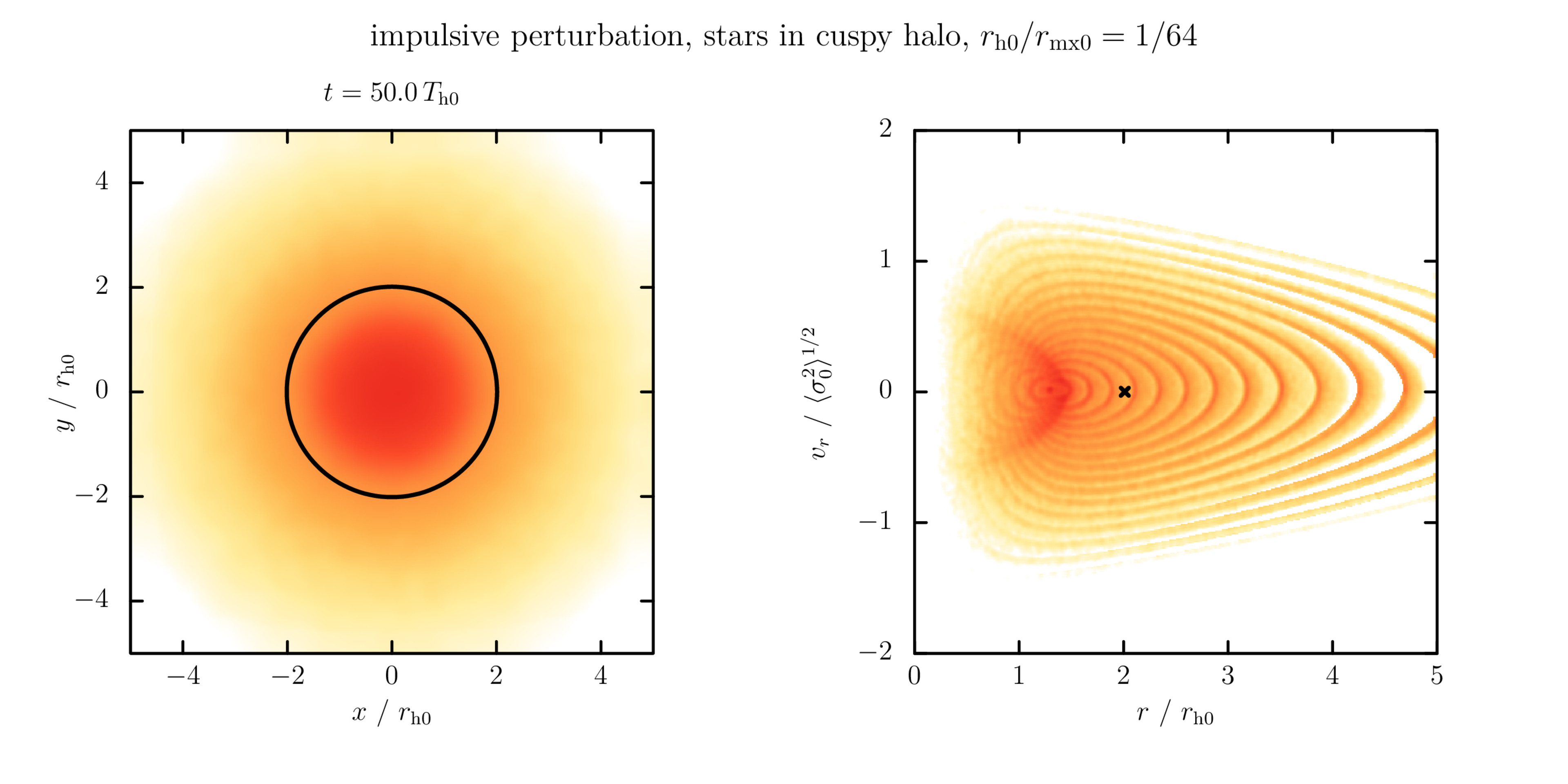}}
 \caption{Animated figure supplementing Fig.~\ref{fig:phasemixing}a. The video shows the time evolution of stellar tracer particles in configuration space $\{x,y\}$ and in phase space $\{ r, v_r\}$ in response to an impulsive perturbation in a cuspy dark matter halo. File size $9.2\,\mathrm{MB}$, dimensions $3840\times1920\,\mathrm{px}^2$, duration $29\,\mathrm{s}$. The animated figure is available in the \href{https://arxiv.org/src/2502.19475v2/anc}{arXiv ancillary files}.}
 \label{fig:ani_14a}
\end{figure}

\begin{figure}[b]
\fbox{\includegraphics[width=8.5cm]{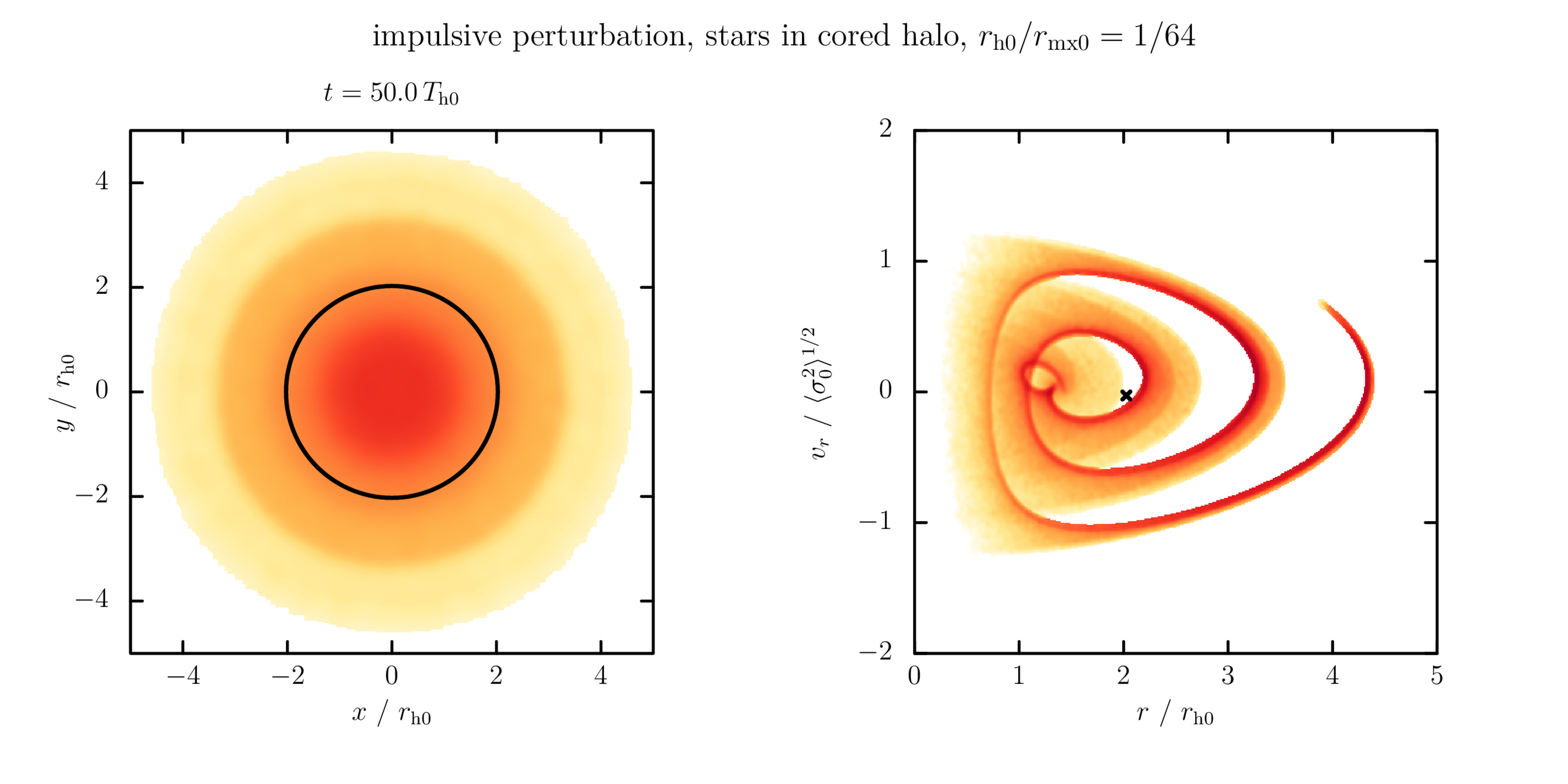}}
 \caption{Animated figure supplementing Fig.~\ref{fig:phasemixing}b. The video shows the time evolution of stellar tracer particles in configuration space $\{x,y\}$ and in phase space $\{ r, v_r\}$ in response to an impulsive perturbation in a cored dark matter halo. File size $8.0\,\mathrm{MB}$, dimensions $3840\times1920\,\mathrm{px}^2$, duration $29\,\mathrm{s}$. The animated figure is available in the \href{https://arxiv.org/src/2502.19475v2/anc}{arXiv ancillary files}.}
 \label{fig:ani_14b}
\end{figure}

\section{Supplementary animated figures}
\label{Appendix:Animations}

This appendix contains brief descriptions of the animated figures \ref{fig:ani_13left}, \ref{fig:ani_13right}, \ref{fig:ani_14a} and \ref{fig:ani_14b}, provided on the journal website.